





\def\marginsA{
    \voffset=0.5truein
    \vsize=8.0truein
    \hoffset=0.truein
    \hsize=6.5truein
    }

\voffset=0.truein
\hoffset=0.truein
\vsize=8.9truein
\hsize=6.5truein
\baselineskip=20.pt
\tolerance=10000
\pretolerance=10000
\hyphenpenalty=10000
\def\ApJstyle{\textstyle}
\def\compact{\baselineskip=12pt \marginsA \def\ApJstyle{\scriptstyle} }

\def\etal{et~al.\ }
\def\oneskip{\vskip\baselineskip}

\def\coverpage{
    \voffset=0.0truein
    \vsize=8.9truein
    \baselineskip=20pt
    \nopagenumbers        }


\def\makeheadline{
\vbox to 0pt{\vskip-40.1pt
  \line{\vbox to 8.5pt{}\the\headline}\vss}
  \nointerlineskip      }

\def\specialpage{\vfill\supereject\vglue .5truein\countspecial=\number\pageno
                 \counteqn=0\countfig=0}
\def\blankfootline{\hfil}
\def\blankheadline{\hfil}
\def\pagenumberfootline{\hss\tenrm\folio\hss}
\def\pagenumberheadline{\hfil\tenrm\folio}
\newcount\countspecial
\newcount\countnopagenumber
\headline={\ifnum\pageno=\countnopagenumber\blankheadline
           \else {\ifnum\pageno=\countspecial\blankheadline
                  \else\pagenumberheadline\fi}
           \fi}
\footline={\ifnum\pageno=\countnopagenumber\blankfootline
           \else {\ifnum\pageno=\countspecial\pagenumberfootline
                  \else\blankfootline\fi}
           \fi}





\newcount\counttemp
\newcount\counteqn
\counteqn=0
\def\eqnnumber#1){\number\counteqn \rm #1)}
\def\eqn#1){\global\advance\counteqn by 1   \eqnnumber \rm #1)}
\def\aeqn#1){\global\advance\counteqn by 1   \eqnnumber \rm #1)}
\def\showeqn#1){\counttemp=\counteqn \advance\counttemp by #1
                \number\counttemp)}
\def\showeqnsq#1]{\counttemp=\counteqn \advance\counttemp by #1
                \number\counttemp]}
\def\showeqnx#1,#2){\counttemp=\counteqn \advance\counttemp by #1
                \number\counttemp#2)}

 
\newcount\counttable
\counttable=0
\def\tablenumber#1.{\number\counttable \rm #1.}
\def\tabler#1.{\global\advance\counttable by 1   \tablenumber \rm #1.}


\newcount\countfig
\countfig=0
\def\fignumber{\number\countfig}
\def\fig{\global\advance\countfig by 1  \fignumber}
\def\showfig#1){\counttemp=\countfig \advance\counttemp by #1
                \number\counttemp)}


\newcount\countsection
\countsection=0
\def\sectionnumber{\number\countsection}
\def\section{\global\advance\countsection by 1  \sectionnumber}
\def\showsection#1){\counttemp=\countsection \advance\counttemp by #1
                \number\counttemp)}


\def\lapprox{\hbox{\lower .8ex\hbox{$\,\buildrel < \over\sim\,$}}}
\def\gapprox{\hbox{\lower .8ex\hbox{$\,\buildrel > \over\sim\,$}}}
\def\gradient{\hbox{\raise .5ex\hbox{$\bigtriangledown$}}}


\def\reference{\parindent=0pt\tolerance=300}
\def\refrepeat{\underbar{\ \ \ \ \ \ \ \ \ \ \ }.\ }
\def\refindent{\par\hangindent=15pt\hangafter=1}
\def\refpaper#11#2,#3,#4,#5.{\refindent#11#2,#3,#4,#5}
\def\refedited#1in#2,#3)#4.{\refindent#1in#2,#3)#4}
\def\refbook#11#2,#3(#4)#5.{\refindent#11#2,#3(#4)#5}

\def\uncatcodespecials{\def\do##1{\catcode\lq##1=12 }\dospecials}
\newcount\lineno 

{\obeyspaces\global\let =\ } 
\def\setupverbatimno{\tt \lineno=0
     \def\par{\leavevmode\endgraf} \catcode`\`=\active
     \obeylines \uncatcodespecials \obeyspaces
     \everypar{\advance\lineno by1 \llap{\sevenrm\the\lineno\ \ }}}
{\obeyspaces\global\let =\ } 



{\obeyspaces\global\let =\ } 


\def\Romannumeral#1.{\uppercase\expandafter{\romannumeral#1}}
\def\hyph{\vrule height 2.35pt width 2.7pt depth -2.10pt} 
{
\catcode`\@=11
\global\def\Biggg#1{{\hbox{$\left#1\vbox to 25.pt{}\right.\n@space$}}}
\global\def\Bigggl{\mathopen\Biggg}
\global\def\Bigggr{\mathclose\Biggg}
}





\compact                     
\baselineskip=12pt           


\def\Xray{X\hyph\hskip 1.pt ray}


\countnopagenumber=\number\pageno
{  
\coverpage

\vglue .5truein
\centerline{A GREY RADIATIVE TRANSFER PROCEDURE}
\centerline{FOR $\gamma$-RAY TRANSFER IN SUPERNOVAE}
            
\vfill
\centerline{DAVID J.~JEFFERY}
\oneskip
\centerline{Department of Physics, University of Nevada, Las Vegas}
\centerline{Las Vegas, Nevada 89154-4002, U.S.A.}
\centerline{email:  jeffery@physics.unlv.edu}

\vfill

\vfill

{
\baselineskip=12pt  
\leftline{40 pages}  
\leftline{1 figure}
\leftline{3 tables}
\leftline{Full address:}
\obeylines
\narrower
           David J.~Jeffery
           Department of Physics
           University of Nevada, Las Vegas
           4505 S. Maryland Parkway
           Las Vegas, Nevada 89154-4002
           U.S.A.
\oneskip
           email:  jeffery@physics.unlv.edu
}

\oneskip

\eject
} 

\specialpage
\pageno=2



\countspecial=\number\pageno
\counteqn=0
\centerline{ABSTRACT}\nobreak
\vbox{                       
\oneskip\nobreak
\narrower
     The $\gamma$-ray transfer in supernovae for the purposes
of energy deposition in the ejecta can be approximated fairly
accurately as frequency-integrated (grey) radiative transfer using
a mean opacity as shown by Swartz, Sutherland, \&~Harkness (SSH).
In SSH's grey radiative transfer procedure (unoptimized)
the mean opacity is a pure absorption opacity and it is a constant
aside from a usually weak composition dependence.
The SSH procedure can be optimized by using a fitted constant mean
absorption opacity for the whole supernova at a given time.
The optimum value of their mean opacity (which depends on the
overall composition and optical depth of the supernova)
is obtained by fitting to more accurate Monte Carlo calculations.
No fitting is needed in the optically thick limit and
(not counting fine adjustments to account for
time-dependent and non-static radiative transfer effects)
in the optically thin limit.

     In this paper, we present a variation on the SSH~procedure which
uses multiple mean opacities which do not need to be obtained by
fitting and which have both absorption and scattering components.
There is a mean opacity for each order of Compton scattering.
(Compton scattering is the dominant form of $\gamma$-ray opacity
in supernovae.)
The zeroth order $\gamma$-ray field (i.e., the direct field from
the nuclear decay) is calculated numerically as in the SSH procedure.
The scattered (i.e., nonzeroth order) $\gamma$-ray fields at a point
are calculated by assuming that 
the scattered $\gamma$-ray source functions
at that point (i.e., local to that point) can be used for the
whole of the ejecta.
This local-state (LS) approximation permits an analytic solution
for the $\gamma$-ray transfer of scattered $\gamma$-ray fields.
The LS~approximation is admittedly crude, but the scattered 
fields are always of lesser importance to the energy deposition. 
Since the LS~approximation is, however, the distinguishing mark of our
procedure, we call our procedure the LS grey radiative transfer
procedure or LS~procedure for short.

     Besides the LS~approximation we also need to approximate
angle-dependent Compton opacity for the analytic solution of
the scattered $\gamma$-ray fields.
We call the approximated Compton opacity the iso-Compton opacity.

     We give only a limited test of the accuracy of our procedure.
For a standard Type~Ia supernova (SN~Ia) model the uncertainty in
$\gamma$-ray energy deposition is estimated to be of order $10\,$\%
or less.
This level of accuracy is often adequate since the deposition is used in 
spectral synthesis calculations which have uncertainties 
of the same order from the atomic data used.

     Since finding the optimum SSH mean opacity requires doing
the detailed (e.g., Monte Carlo) radiative transfer one wants to avoid in
using a simplified $\gamma$-ray energy deposition procedure,
the LS~procedure may be the best choice for that simplified
procedure.
The extra effort in developing and running an LS~procedure code
beyond that of an SSH procedure code is small.

     The LS~procedure code used for this paper can be obtained by request
from the author.
This code (excluding comment lines, auxiliary subroutines,
and data statements) is about 260 lines long.

     For completeness and easy reference, we include in this paper
a review of the $\gamma$-ray opacities important in supernovae, a
discussion of the appropriate mean opacity prescription, and a discussion of
the errors arising from neglecting time-dependent and non-static
radiative transfer effects.
\vskip 2\baselineskip
}                            
\vfill\eject

\specialpage

\centerline{1.\ \ INTRODUCTION}\nobreak
\vskip\baselineskip\nobreak
     The decay of radioactive elements synthesized in 
supernova explosions is one of the most important sources of the
energy driving observable supernova luminosity.
For Type~Ia supernovae (SNe~Ia)
(e.g., Colgate, Petschek, \&~Kriese 1980;   
Harkness 1991) and probably Type~Ic supernovae (e.g.,
Young, Baron, \&~Branch 1995),              
almost all the observed luminosity comes
from radioactive decay.
For the other supernova types, radioactive decay
drives much, and probably in most cases most, of the observable
luminosity in the nebular epoch or even earlier as suggested
by Type~II supernova SN~1987A (e.g., Woosley 1988) 
and Type~IIb supernova SN~1993J (e.g., Young~\etal 1995). 
The overwhelmingly dominant decay chain for the observable
epoch of most supernovae is
$^{\ApJstyle 56}$$\rm Ni\to^{\ApJstyle 56}$$\rm Co\to^{\ApJstyle 56}$$\rm Fe$
with \hbox{half-lives} of 5.9 and~77.27~days for the first and second
decays, respectively (Huo 1992).
The first decay releases almost all its energy in the form of
$\gamma$-rays and the second in $\gamma$-rays and, in $19\,$\% of
the decays, in positrons (Browne \&~Firestone 1986;  Huo 1992).
The $\gamma$-ray and mean positron kinetic energy are in the
range $\sim 0.15$--$3.6\,$MeV.  
The longer lived radioactive species $^{\ApJstyle 57}$Co and
$^{\ApJstyle 44}$Ti are likely to become important only after
about day~800 after explosion as suggested by the observations
of SN~1987A (e.g., Nomoto~\etal 1994, p.~546ff) and tests we
have done for SNe~Ia.  
Very few supernovae are observed so late and by then other
energy sources such as circumstellar interaction or pulsar remnants may
have become important also (e.g., Fransson 1994, p.~731ff), except
probably for SNe~Ia.

     After earliest and usually unobserved times, the $\gamma$-rays
and positrons deposit energy in a 
relatively cold ($T\lapprox 10^{\ApJstyle 4}\,$K), low-ionization
state or nearly neutral supernova medium.
This deposited energy is effectively mainly in the form of
fast electrons from Compton scattering, photoionization,
and positron collisions.  
The fast electrons ionize and excite atoms and heat the
gas through electron collisions.
The fast electron energy gets transformed into other forms
of energy by a complicated cascade process that is described by, e.g.,
Fransson (1994, p.~688ff) and Liu \&~Victor (1994).

     The dispersion of the decay energy through the supernova
is by $\gamma$-ray radiative transfer and possibly by the
positrons and the fastest fast electrons which both lose kinetic
energy at an increasing rate as they are slowed down.
The simplest assumption is that the positrons and fast electrons
do not disperse their kinetic energy, but only deposit it locally.
This would certainly be the case if there are tangled magnetic fields
of even weak strength in the supernovae (e.g., Colgate~\etal 1980; 
Chan \&~Lingenfelter 1993).  
Unfortunately, little is certain about the magnetic fields in
supernovae and it may be that they are radially combed out
in which case considerable energy transport via positrons and
some of the fast electrons may occur.
It is at least certain that the positrons must be much more trapped
than the $\gamma$-rays since theoretical SN~Ia light curves and
absolute spectra at 
late times require positron kinetic energy deposition to dominate
$\gamma$-ray energy deposition in order to match observations
(e.g., Liu, Jeffery, \&~Schultz 1997a, b).
Positron transport may be an important process and it is being
actively investigated (e.g., Colgate~\etal 1980;  Chan \&~Lingenfelter
1993;  Ruiz-Lapuente 1997;  Milne, The, \&~Leising 1997;
Ruiz-Lapuente \&~Spruit 1998), but it is outside of the scope of
the present paper.

     We note that the positrons (unless they escape the supernova
altogether) will annihilate with electrons either by 
a three-continuum-photon process from a triplet
positronium state
or by a two-$m_{\ApJstyle e}c^{\ApJstyle 2}$-photon process
from a singlet positronium state or directly with free or
bound electrons (e.g., Brown \&~Leventhal 1987). 
The positrons probably lose most of their kinetic energy
before annihilation---it effectively goes into fast electron
energy---and their rest mass energy and the rest mass energy
of the annihilated electrons become
part of the $\gamma$-ray flux. 
We assume in this paper that the positrons are completely locally
trapped
so that the annihilation $\gamma$-rays can be treated on the
same footing as the $\gamma$-rays coming directly from the radioactive
decay.

      The $\gamma$-ray transfer, which is the topic of the present
paper, can be treated to high accuracy by Monte Carlo calculations in
which all important physical processes can be handled in detail.
For the calculation of the spectra of $\gamma$-rays that escape the ejecta
treatments at least as detailed as Monte Carlos are probably necessary.
Such Monte Carlo spectrum calculations have been done by, e.g.,
Ambwani \&~Sutherland (1988),
H\"oflich, Khokhlov, \&~M\"uller (1992),
Kumagai~\etal (1993), Ruiz-Lapuente~\etal (1993), G\'omez-Gomar,
Isern, \&~Jean (1998)
and 
H\"oflich, Wheeler, \&~Khokhlov (1998).
Another high accuracy approach to $\gamma$-ray transfer would be to use
the comoving frame frame formalism including all special relativistic
effects (Mihalas 1980a) and angle and frequency partial redistribution
effects (Mihalas 1980b).
A possible third approach would be to do an observer frame calculation
including all special relativistic and angle and frequency
partial redistribution effects.
In this third approach, the source functions for scattered
$\gamma$-ray fields could be obtained by a $\Lambda$-iteration
(e.g., Mihalas 1978, p. 147ff).
In supernova $\gamma$-ray transfer case, the $\Lambda$-iteration
is expected to converge since the zeroth order field source functions
are specified and scattering is only locally important until the
ejecta is optically thin.
Like the Monte Carlo approach, this third approach could be generalized
to three-dimensional $\gamma$-ray transfer.
Time-dependent effects could probably be handled
in the third approach as well.

     The detailed accurate $\gamma$-ray transfer procedures
(which can be computationally intensive and/or difficult to
code) are not, however, needed to obtain fairly accurately the
$\gamma$-ray energy deposited in the ejecta in fast electron energy.
A simple frequency-integrated (grey) radiative transfer procedure 
can do this
(Colgate~\etal 1980;
Sutherland \&~Wheeler 1984;
Ambwani \&~Sutherland 1988;
Swartz, Sutherland, \&~Harkness 1995, hereafter SSH).
In such procedures, a
mean opacity replaces the frequency-specific opacity
needed in detailed (frequency-dependent) radiative transfer procedures. 
The basic fortran code for the grey procedures is straightforward to write
and requires no more than of order 100 lines (Sutherland 1996). 
A grey radiative transfer calculation typically requires of order
$10^{\ApJstyle -5}$ of the computational time of a Monte Carlo
calculation (SSH)
and the result, the amount of energy deposited, is actually very
useful since the fast electron cascade process is quite insensitive
to the spectrum of the primary
fast electrons (e.g., Liu \&~Victor 1994).  
Given the $\gamma$-ray energy deposition
and the deposition of the positron kinetic energy, the 
thermal state of the ejecta and its ultraviolet-optical-infrared
(UV-optical-IR) emission can be calculated (e.g., Axelrod 1980;
Ruiz-Lapuente 1997;  Liu~\etal 1997a, b, c).

     A key point for grey radiative transfer calculations is the
choice of the mean opacity.
In earlier work (Colgate~\etal 1980;  Sutherland \&~Wheeler 1984;
Ambwani \&~Sutherland 1988), the mean opacity was 
a pure absorption opacity and a single mean opacity
value was suggested for all locations in supernova ejecta and all
supernova epochs. 
In the unoptimized version of the grey $\gamma$-ray transfer
procedure of SSH (hereafter the SSH procedure),
the mean opacity is a pure absorption opacity and it is a constant
aside from a usually weak composition dependence.
For this opacity SSH suggest 
$\kappa=0.06/\mu_{\ApJstyle e}\,{\rm cm^{\ApJstyle 2}\,g^{\ApJstyle -1}}$
for most cases.
The mean atomic mass per electron $\mu_{\ApJstyle e}$ accounts for the
main composition dependence (see \S~2):  for hydrogen dominated matter,
$\mu_{\ApJstyle e}$ is about 1;  for metal dominated matter, it is about 2.

     The SSH procedure can be optimized by using a fitted constant mean
absorption opacity for the whole supernova at a given time.
This mean opacity value (which depends on the
overall composition and optical depth of the supernova)
is obtained by fitting to more accurate Monte Carlo calculations.
No fitting is needed in the optically thick limit and
(not counting fine adjustments to account for
time-dependent and non-static radiative transfer effects)
in the optically thin limit.

     In the SSH procedure, the actual $\gamma$-ray transfer
is done by a numerical integration solution of the radiative transfer equation. 
For the supernova model examined by SSH (SN~Ia model~W7
[Thielemann, Nomoto, \&~Yokoi 1986]), the optimized SSH~procedure 
obtained an accuracy in energy deposition
of a few percent locally          
and $2\,$\% globally.             

     The level of accuracy obtained by optimized and also the
unoptimized SSH~procedure
is often adequate, particularly for preliminary calculations,
since the energy deposition is used in
spectral synthesis calculations which have uncertainties 
of the same order (e.g., $\sim 10\,$\%) from the atomic data used.
Published calculations using the unoptimized SSH procedure
have been done by, e.g., Houck \&~Fransson (1996) and
Liu~\etal (1997a, b, c).

     It is probable that accuracy in unoptimized SSH procedure
calculations close to that from
optimized ones can be obtained using mean opacity values that
have been estimated based on past Monte Carlo calculations (see \S~6).
Nevertheless, it would be more satisfactory to have a procedure 
which would guarantee reasonable results without
adjusting a free parameter and which would yield further physical insight.
In this paper we present a variation on the SSH~procedure that
does this and that uses multiple mean opacities with both
absorption and scattering components.
This procedure adapts to optical depth conditions, and so no fitting is
required.
(The composition of the supernova model is known {\it a~priori,} and so no
fitting for composition dependence is needed either.)

     In our procedure, there is a mean opacity for each order of
Compton scattering.
(Compton scattering is the dominant form of $\gamma$-ray opacity
in supernovae:  see \S~2).
The zeroth order $\gamma$-ray field (i.e., the direct field from
the nuclear decay) is calculated numerically as in the SSH procedure.
The scattered (nonzeroth order) $\gamma$-ray fields at a point
are calculated by
assuming that the scattered $\gamma$-ray source functions
at that point (i.e., local to that point) can be used for the
whole of the ejecta.
This local-state (LS) approximation permits an analytic solution
for the $\gamma$-ray transfer of scattered $\gamma$-ray fields.
The LS~approximation is admittedly crude, but the scattered 
fields are always of lesser importance to the energy deposition.
Since the LS~approximation, however, is the distinguishing mark of our
procedure, we call our procedure the LS grey radiative transfer
procedure or LS~procedure for short.
A brief presentation of the LS~procedure is given by
Jeffery (1998).

     The LS~procedure, like the SSH~procedure, treats the
$\gamma$-ray transfer as occurring in a time-independent,
static medium.
Note, however, that in the SSH~procedure time-dependent
and non-static effects can be absorbed into the fitting of the mean
opacity and SSH, in fact, do this at least partially
(Sutherland 1998).

     In \S~2 of this paper, we describe the $\gamma$-ray opacities
relevant to supernovae and introduce what we call the iso-Compton
opacity approximation.
In \S~3, we obtain the mean opacity prescription and mean
opacities that we use.
The LS~procedure is presented in \S~4.
Section~5 reviews some of the material needed for the 
treatment of the radioactive sources of energy in supernovae.
In \S~6, we discuss the adequacy of some of the approximations
we make and compare the LS~procedure to the SSH~procedure.
Conclusions are given in \S~7.
In Appendix~A, we discuss the errors arising from the
neglect in the LS~procedure of the effects of time-dependent
and non-static radiative transfer.
Appendix~B proves some of the mathematical properties
of the LS approximation series that we introduce in \S~4.
\vskip 2\baselineskip

\centerline{2.\ \  $\gamma$-RAY OPACITIES IN SUPERNOVAE}\nobreak
\vskip\baselineskip\nobreak
     To begin we should specify our use of the term opacity.
There two common usages.
The first usage is for the inverse of the mean free path;  this
quantity is also called the extinction (e.g., Mihalas 1978, p.~607).
The second usage, which we adopt here, is the extinction 
divided by density.
This usage is much more convenient when discussing $\gamma$-ray
transfer in supernovae since in this case this kind of opacity
is almost entirely
independent of density and depends almost entirely on composition.
In supernovae, density varies with location and time by many orders
of magnitude.

     For the case of supernova $\gamma$-rays,
the opacity can be divided      
into absorption and scattering components.
The former treats the transformation of $\gamma$-ray energy into
some other form which for our case is effectively fast electron
kinetic energy.
The scattering component treats the transformation of $\gamma$-rays into
other $\gamma$-rays.
The sum of the absorption and scattering opacities is the total
opacity or simply the opacity.
Throughout this paper we will use the superscript $R$ as a variable
that replaces a symbol designating a quantity as related to
total (blank), absorption (``a''), or scattering (``s'') opacity:
e.g., the general symbol for opacity (second usage)
$\kappa^{\ApJstyle R}$ stands for $\kappa$, $\kappa^{\ApJstyle\rm a}$,
or $\kappa^{\ApJstyle\rm s}$.
The relation between opacity for a particular particle
and the particle cross section $\sigma^{\ApJstyle R}$ is given by
$$   \kappa^{\ApJstyle R}
       ={n\over\rho}\sigma^{\ApJstyle R} \,\,  ,     \eqno(\eqn)$$
where $n$ is the particle density and $\rho$ is (mass) density.

     In the energy range $0.05$--$50\,$MeV, $\gamma$-rays interact with
matter principally through three processes:
(1)~pair production in the Coulomb field of a nucleus or an electron,
(2)~the
photoelectric effect with bound electrons (which is just $\gamma$-ray
photoionization of an atom or ion), and
(3)~Compton scattering off electrons (e.g., Davisson 1965, p.~37).
Almost all the $\gamma$-rays from the
$^{\ApJstyle 56}$$\rm Ni\to^{\ApJstyle 56}$$\rm Co\to^{\ApJstyle 56}$$\rm Fe$
decay chain and other decay chains important in supernovae
lie in the energy range $0.05$--$50\,$MeV and no $\gamma$-rays
exceed $\sim 3.6\,$MeV in fact (Browne \&~Firestone 1986;  Huo 1992).
Thus the three mentioned processes determine $\gamma$-ray opacity in  
supernovae (see also SSH's Fig.~1).

     For the pair production opacity, one can use
$$ \kappa_{\ApJstyle\rm pair}
={\sigma_{\ApJstyle\rm pair}^{\ApJstyle *}\over m_{\ApJstyle\rm amu}}
\sum_{\ApJstyle i} {X_{\ApJstyle i}Z_{\ApJstyle i}^{\ApJstyle 2}
                    \over A_{\ApJstyle i} }  \,\, , \eqno(\eqn)$$
where the sum is over all the elements, $X_{\ApJstyle i}$ is the mass
fraction
an element, $A_{\ApJstyle i}$ is the element's atomic mass,
$Z_{\ApJstyle i}$ is the nuclear
charge of the element, $m_{\ApJstyle\rm amu}$ is the atomic mass unit, 
and $\sigma_{\ApJstyle\rm pair}^{\ApJstyle *}$ is the atomic
pair production cross section divided by $Z_{\ApJstyle i}^{\ApJstyle 2}$.
An expression for $\sigma_{\ApJstyle\rm pair}^{\ApJstyle *}$ 
(adapted from Hubbell 1969) is
$$ \sigma_{\ApJstyle\rm pair}^{\ApJstyle *}=10^{\ApJstyle -27}\times
\cases{ 
   0 \,\, ,    & $E<2m_{\ApJstyle e}c^{\ApJstyle 2}$; \cr
   0.10063\times\left(E-2m_{\ApJstyle e}c^{\ApJstyle 2}\right) \,\, ,
             & $2m_{\ApJstyle e}c^{\ApJstyle 2}\leq E < 1.5\,$MeV; \cr
   \left[0.0481+0.301\times\left(E-1.5\right)\right] \,\, ,
             & $E \geq 1.5\,$MeV, \cr}  \eqno(\eqn)$$
where $\sigma_{\ApJstyle\rm pair}^{\ApJstyle *}$ is in cm$^{\ApJstyle 2}$,
$m_{\ApJstyle e}$ is the electron mass,
and $E$ is the $\gamma$-ray energy measured in MeV.
(Note the constant of the third case of equation~(\showeqn -0)
could be changed 0.048101329 to ensure better continuity for
$\sigma_{\ApJstyle\rm pair}^{\ApJstyle *}$ although there is no
change in physical accuracy.)
We note that the positron created in pair production will annihilate
to form $\gamma$-rays (see also \S~1).
Thus in an effective sense, especially with assumption of
time independence,
pair production opacity can be regarded as having a scattering component.
We assume that the positron loses all of its kinetic energy
before annihilation, and thus the pair production absorption 
opacity is given by
$$ \kappa_{\ApJstyle\rm pair}^{\ApJstyle\rm a}=
    \kappa_{\ApJstyle\rm pair}
     \left({E-2m_{\ApJstyle e}c^{\ApJstyle 2}\over E}\right) \eqno(\eqn)$$
and the pair production scattering opacity by
$$ \kappa_{\ApJstyle\rm pair}^{\ApJstyle\rm s}=
    \kappa_{\ApJstyle\rm pair}{2m_{\ApJstyle e}c^{\ApJstyle 2}\over E}
                                                \,\, .     \eqno(\eqn)$$ 
For the development of the LS~procedure formalism (see \S~3.2), we assume 
that a pair production scattering always results in
two $m_{\ApJstyle e}c^{\ApJstyle 2}$ $\gamma$-rays and not in three
$\gamma$-rays with a continuum of energies.
This assumption introduces negligible error because it turns out that
pair production opacity is of very small importance (see \S~3.2)
The angular redistribution of pair production scattering
is probably very isotropic because of the complicated path the
positron will take in slowing down if for no other reason.
We assume that it is completely isotropic.

     We take the photoelectric opacity to be entirely absorption opacity:
i.e., we assume any low energy X-rays resulting from a photoelectric
effect ionized and excited atom will be locally absorbed eventually
into fast electron energy or the local thermal pool.
The photoelectric opacity can
be approximated quite accurately in the range $0.01$--$1\,$MeV by
$$ \kappa_{\ApJstyle\rm pe}
=\kappa_{\ApJstyle\rm pe}^{\ApJstyle *}
\left({E\over 0.1\,{\rm MeV}}\right)^{\ApJstyle -3} \,\, , \eqno(\eqn)$$
where
$$ \kappa_{\ApJstyle\rm pe}^{\ApJstyle *}=
{1\over m_{\ApJstyle\rm amu}}\left(\sum_{\ApJstyle i}
  {X_{\ApJstyle i}\over A_{\ApJstyle i}}
    \sigma_{\ApJstyle {\rm pe,} i}^{\ApJstyle 0.1}
       \right) \eqno(\eqn)$$
(e.g., SSH).
The $\sigma_{\ApJstyle {\rm pe,} i}^{\ApJstyle 0.1}$ values are the
photoelectric cross sections of the atoms at $0.1\,$MeV.
To be more accurate, one can construct tables of $\kappa_{\ApJstyle\rm pe}$
as a function of energy for different compositions.
The cross section data needed to construct
$\kappa_{\ApJstyle\rm pe}^{\ApJstyle *}$ values or 
$\kappa_{\ApJstyle\rm pe}$ tables can be found
in, e.g., Veigele (1973).

     The Compton opacity is given to a good approximation by
$$ \kappa_{\ApJstyle\rm C}^{\ApJstyle R}=
{n_{\ApJstyle e}^{\ApJstyle\rm total}\over\rho}
 \sigma_{\ApJstyle\rm C}^{\ApJstyle R}
={\sigma_{\ApJstyle\rm C}^{\ApJstyle R}
  \over m_{\ApJstyle\rm amu}\mu_{\ApJstyle e}}
                                                  \,\, ,       \eqno(\eqn)$$
where $n_{\ApJstyle e}^{\ApJstyle\rm total}$ is the total electron density
counting both free and bound electrons,
$\sigma_{\ApJstyle\rm C}^{\ApJstyle R}$ is the
Compton cross section, and
$\mu_{\ApJstyle e}$ is the mean atomic mass per electron.
The expression for $\mu_{\ApJstyle e}$ is
$$ \mu_{\ApJstyle e}^{\ApJstyle -1}=
  \sum_{\ApJstyle i}{X_{\ApJstyle i}Z_{\ApJstyle i}\over A_{\ApJstyle i}}
                                                  \,\, ,       \eqno(\eqn)$$
where the sum is again over all elements.
The $Z_{\ApJstyle i}$ is again the nuclear charge
since we make the assumption that all electrons, free or bound, act
as if they were free.
This is a good assumption for $\gamma$-rays with energies much
larger electron binding energies (e.g., Davisson 1965, p.~49).
The fact that bound electrons are spatially concentrated about atoms
makes no difference.
The effect of any one electron is minute, and so
the effects of all the electrons in an atom just add linearly.
In the supernova case, the $\gamma$-rays after several
scatterings have lost most of their energy, but
nevertheless still mostly have energies much larger than electron 
binding energies.
(The electron binding energies for important atoms are
$\lapprox0.01\,$MeV [e.g., Veigele 1973].
For mean $\gamma$-ray energies for the first 5 orders of scattering
[as our approximate treatment gives them]
see \S~3.2, Tables~I and~II.
Recall also that the supernova medium we consider is in a low-ionization
or nearly-neutral state, and so the problem of tightly bound
electrons in highly-charged ions does not arise.) 
Thus the error in assuming all electrons act as if they were free is small.
Additionally, in the metal-rich compositions such as those of
SNe~Ia and the deep interior of the other supernova
types, the photoelectric opacity can dominate for energies
below $\sim 0.1\,$MeV (see SSH's Fig.~1), and thus in these cases
the error in using Compton opacity for bound electrons is smaller
still. 

     Below we give the Compton opacity formulae which have been
adapted from Davisson (1965, p.~51ff).
Note that Compton opacity has both absorption and scattering
components since Compton scattering is not coherent
(i.e., not elastic or energy-conserving).
The Compton total cross section is given by
$$ \sigma_{\ApJstyle\rm C}=
   \sigma_{\ApJstyle e}\left({3\over4}\right)\left\{
\left({1+\alpha\over\alpha^{\ApJstyle 2}}\right)
 \left[{2\left(1+\alpha\right)\over 1+2\alpha}
       -{\ln\left(1+2\alpha\right)\over \alpha}\right]
 +{\ln\left(1+2\alpha\right)\over 2\alpha}
 -{1+3\alpha\over\left(1+2\alpha\right)^{\ApJstyle 2}} \right\}
                                \,\, , \eqno(\eqn)$$
where
$$\sigma_{\ApJstyle e}=0.66524616(18)\times10^{\ApJstyle -24}
                        \,{\rm cm^{\ApJstyle 2}}       \eqno(\eqn)$$
(Cohen \&~Taylor 1987) is the Thomson cross section (with uncertainty
in the last digits in the brackets) and
$$ \alpha={E\over m_{\ApJstyle e}c^{\ApJstyle 2}} \eqno(\eqn)$$
is the $\gamma$-ray energy in units of the electron rest energy.
The Compton absorption cross section is given by
$$\eqalignno{
 \sigma_{\ApJstyle\rm C}^{\ApJstyle\rm a}&=
   \sigma_{\ApJstyle e}\left({3\over8}\right)\Bigggl[
\left({-3-2\alpha+\alpha^{\ApJstyle 2}\over \alpha^{\ApJstyle 3}}\right)
  \ln\left(1+2\alpha\right) \cr
&\qquad\qquad\qquad
+{2\left(9+51\alpha+93\alpha^{\ApJstyle 2}+51\alpha^{\ApJstyle 3}
          -10\alpha^{\ApJstyle 4}\right)
  \over 3\alpha^{\ApJstyle 2}\left(1+2\alpha\right)^{\ApJstyle 3} }\Bigggr]
                                                 &(\eqn) \cr}$$
and the Compton scattering cross section by
$$ \sigma_{\ApJstyle\rm C}^{\ApJstyle\rm s}=
   \sigma_{\ApJstyle e}\left({3\over8}\right)\left[
{\ln\left(1+2\alpha\right)\over\alpha^{\ApJstyle 3}}
-{2\left(1+\alpha\right)\left(1+2\alpha-2\alpha^{\ApJstyle 2}\right)
  \over \alpha^{\ApJstyle 2}\left(1+2\alpha\right)^{\ApJstyle 2} }
+{ 8\alpha^{\ApJstyle 2}\over 3\left(1+2\alpha\right)^{\ApJstyle 3} } \right]
                                          \,\, .              \eqno(\eqn)$$
Equations~(\showeqn -1) and~(\showeqn -0) are angle-averaged expressions
since Compton scattering is anisotropic and the energy loss on scattering
is angle-dependent.
To second order in $\alpha$, the cross section expressions are
$$\eqalignno{
 \sigma_{\ApJstyle\rm C,2nd}&=
   \sigma_{\ApJstyle e}\left(1-2\alpha+{26\over5}\alpha^{\ApJstyle 2}\right)
                                         \,\,  ,  &(\eqn)\cr
 \sigma_{\ApJstyle\rm C,2nd}^{\ApJstyle\rm a}&=
   \sigma_{\ApJstyle e}\left(\alpha-{21\over5}\alpha^{\ApJstyle 2}\right)
                                         \,\,  ,  &(\eqn)\cr
\noalign{\noindent\rm and}
 \sigma_{\ApJstyle\rm C,2nd}^{\ApJstyle\rm s}&=
   \sigma_{\ApJstyle e}\left(1-3\alpha+{47\over5}\alpha^{\ApJstyle 2}\right)
                                         \,\,  .  &(\eqn)\cr }$$
We see that in the limit of $\alpha$ going to 0 Compton scattering
reduces to coherent Thomson scattering.
The asymptotic forms of the cross section expressions as $\alpha$ goes
to infinity are
$$\eqalignno{
 \sigma_{\ApJstyle\rm C,asy}&=
   \sigma_{\ApJstyle e}\left({3\over 8\alpha}\right)
    \left[\ln(\alpha)+\ln(2)+{1\over2}\right]
                                         \,\,  ,  &(\eqn)\cr
 \sigma_{\ApJstyle\rm C,asy}^{\ApJstyle\rm a}&=
   \sigma_{\ApJstyle e}\left({3\over 8\alpha}\right)
    \left[\ln(\alpha)+\ln(2)-{5\over6}\right] \,\, ,  &(\eqn)\cr
\noalign{\noindent\rm and}
 \sigma_{\ApJstyle\rm C,asy}^{\ApJstyle\rm s}&=
   \sigma_{\ApJstyle e}\left({3\over 8\alpha}\right)
    \left({4\over3}\right) 
                                         \,\,  .  &(\eqn)\cr }$$

      The Compton total cross section decreases with $\alpha$ from
$\sigma_{\ApJstyle e}$
at $\alpha=0$ to 0 at $\alpha=\infty$. 
The only stationary point is the minimum at infinity.
The Compton scattering cross section has the same behavior, except that it
decreases more rapidly.
The Compton absorption cross section rises from 0 at $\alpha=0$ to
a maximum of $\sim 0.14838408\sigma_{\ApJstyle e}$ at
$\alpha\approx0.98212734$ (i.e., an energy of $\sim 0.50186615\,$MeV)
and then decreases to 0 at $\alpha=\infty$.
Aside from the maximum, the only stationary point is the minimum at infinity.
The fractional scattering opacity,
$\kappa_{\ApJstyle\rm C}^{\ApJstyle\rm s}/\kappa_{\ApJstyle\rm C}
=\sigma_{\ApJstyle\rm C}^{\ApJstyle\rm s}/\sigma_{\ApJstyle\rm C}$,
decreases for all $\alpha$.
It is 1 at $\alpha=0$ and goes asymptotically to 0 at $\alpha=\infty$;
the only stationary point is the minimum at $\alpha=\infty$.
The fractional absorption opacity,
$\kappa_{\ApJstyle\rm C}^{\ApJstyle\rm a}/\kappa_{\ApJstyle\rm C}
=\sigma_{\ApJstyle\rm C}^{\ApJstyle\rm a}/\sigma_{\ApJstyle\rm C}$,
behaves, of course, in a complementary manner to the fractional scattering
opacity:  it increases for all $\alpha$, is 0 at $\alpha=0$,
and goes to 1 at $\alpha=\infty$ which is the only stationary point.

     Compton opacity is, as mentioned in \S~1, the dominant opacity
for supernova $\gamma$-rays.
In the metal-rich composition of SNe~Ia, in which
they are strongest,
pair production opacity only begins to be important above $\sim 3\,$MeV
and only dominates at $\sim 10\,$MeV and photoelectric opacity only begins
to be important below $\sim 0.3\,$MeV and only dominates at $\sim 0.1\,$MeV
(see SSH's Fig.~1).
Since very roughly speaking the unscattered decay $\gamma$-rays in
supernovae have energies of order $1\,$MeV and
$1\,$MeV $\gamma$-rays lose about half their energy in Compton
scattering,  
it is clear in SNe~Ia that most $\gamma$-ray energy must be lost in Compton
scattering.
In other kinds of supernovae where metallicity is lower, Compton
opacity is even more important.

     A key point about Compton scattering is that its angular redistribution
is forward peaked and the degree of forward peaking increases with
increasing $\gamma$-ray energy.
The ratio of the forward to the backward scattering differential
cross sections (for energy, not photon number) is
$$ {\left(1+2\alpha\right)^{\ApJstyle 3}\over
   \displaystyle \left(1 +{2\alpha^{\ApJstyle 2}\over 1+2\alpha}\right) }
                                                           \,\, . \eqno(\eqn)$$
For 1$\,$MeV photons, this ratio is $\sim 46.372391$.
The angle-dependent energy reduction factor on Compton scattering
is given by
$$ {1 \over 1+\alpha\left(1-\cos\theta\right) } \,\, , \eqno(\eqn)$$
where $\theta$ is the scattering angle
(e.g., Davisson 1965, p.~50).
From equation~(\showeqn -0), it follows that forward scattered photons
lose no energy at all.

     It is clear that a substantial fraction of Compton scattering is
nearly-forward and nearly-coherent.
This fraction can almost be neglected since it barely affects the
$\gamma$-ray flux. 
On the other hand the non-forward, noncoherent scattering fraction 
of Compton scattering is relatively small.
One concludes that the total scattering component of Compton opacity 
is of relatively low significance for $\gamma$-ray transfer. 
Therefore, one could try to approximate Compton opacity by
using only its absorption component and neglecting the scattering
component.
To see how this would work consider a medium with an
opacity with both absorption and scattering components.
One now does the radiative transfer through the medium
neglecting the scattering component.
This absorption-only approximation will tend to underestimate
absorption in cases of finite, nonzero optical thickness. 
The scattering component (the non-forward scattering component
to be precise) of the opacity tends to increase the
trapping of flux in the medium by randomizing its direction and
the trapped flux has more opportunities to be absorbed. 
Without the scattering component some absorption tends
to be missed. 
Of course, if the medium is in the optically thick limit,
scattering will not add to the trapping and absorption,
and the absorption-only approximation will work well.
On the other hand, the absorption-only approximation
also gives exactly the right absorption in the optically
thin limit where a $\gamma$-ray scatters on average much
less than once.

     For the LS~procedure we wish to exploit the low significance
of the scattering component of Compton opacity, but without
making the simple absorption-only approximation.
We will do this using two approximations:
(1) an approximation to Compton
opacity that we call the iso-Compton opacity
and
(2) an approximate treatment (that uses the LS~approximation)
of the non-forward, noncoherently scattered flux.
The second approximation we describe in \S~4.
To make the iso-Compton opacity approximation we separate
the Compton opacity into two approximate components:
an isotropic, noncoherently scattering component 
(the iso-Compton component)
and a forward, coherently scattering component
(the forward component).
The iso-Compton component is the iso-Compton opacity itself.
Since the forward component is pure scattering its total
and scattering cross sections are equal and its absorption
cross section is zero. 

    The forward component is effectively a zero opacity and simply
does not appear in the radiative transfer calculations.
The iso-Compton cross sections 
are obtained by subtracting the forward component cross sections
from the corresponding Compton cross sections:  i.e.,
$$ \sigma_{\ApJstyle\rm C}({\rm iso})
      =\sigma_{\ApJstyle\rm C}
       - \sigma_{\ApJstyle\rm C}^{\ApJstyle\rm s}
           \left(\theta_{\ApJstyle\rm f}\right) 
                                               \,\, ,  \eqno(\eqn)$$
$$ \sigma_{\ApJstyle\rm C}^{\ApJstyle\rm a}({\rm iso})
      =\sigma_{\ApJstyle\rm C}^{\ApJstyle\rm a}-0
      =\sigma_{\ApJstyle\rm C}^{\ApJstyle\rm a}
                                               \,\, ,  \eqno(\eqn)$$
and
$$ \sigma_{\ApJstyle\rm C}^{\ApJstyle\rm s}({\rm iso})
      =\sigma_{\ApJstyle\rm C}^{\ApJstyle\rm s}
       -\sigma_{\ApJstyle\rm C}^{\ApJstyle\rm s}
          \left(\theta_{\ApJstyle\rm f}\right)
                                                  \,\, ,\eqno(\eqn)$$
where ``iso'' stands for iso-Compton opacity and 
$\sigma_{\ApJstyle\rm C}^{\ApJstyle\rm s}
\left(\theta_{\ApJstyle\rm f}\right)$ is the forward component.
To obtain the iso-Compton component opacities we just replace the
Compton cross sections in equation~(8) by the iso-Compton cross
sections.
We treat the energy reduction on iso-Compton scattering to be a
constant for all angle.
For the energy reduction factor we use 
$\sigma_{\ApJstyle\rm C}^{\ApJstyle\rm s}({\rm iso})
 /\sigma_{\ApJstyle\rm C}({\rm iso})$.
The energy reduction factor of the forward component is,
of course, 1.

     The physical picture of the forward component cross
section is that it is the Compton scattering cross section 
$\sigma_{\ApJstyle\rm C}^{\ApJstyle\rm s}
  \left(\theta_{\ApJstyle\rm f}\right)$
for a cone of scattering directions
with opening angle $\theta_{\ApJstyle\rm f}$ 
centered on the forward direction that has been reassigned to the
forward direction.
The Compton absorption opacity for the cone has been reassigned to the
iso-Compton component. 
The cone's average energy reduction factor is
$\sigma_{\ApJstyle\rm C}^{\ApJstyle\rm s}
  \left(\theta_{\ApJstyle\rm f}\right)
/\sigma_{\ApJstyle\rm C}
  \left(\theta_{\ApJstyle\rm f}\right)$,
where $\sigma_{\ApJstyle\rm C}\left(\theta_{\ApJstyle\rm f}\right)$
is the Compton total cross section for the cone.
The physical picture of the iso-Compton component is that
it is the opacity for other directions plus the absorption
component opacity from the cone
which have been spread uniformly over the whole scattering sphere.
The average energy reduction factor for the other
directions is
$\left[\sigma_{\ApJstyle\rm C}^{\ApJstyle\rm s}-
\sigma_{\ApJstyle\rm C}^{\ApJstyle\rm s}
  \left(\theta_{\ApJstyle\rm f}\right)\right]
/\left[\sigma_{\ApJstyle\rm C}-
  \sigma_{\ApJstyle\rm C}
  \left(\theta_{\ApJstyle\rm f}\right)\right]$
and this is larger than energy reduction factor assigned to the
iso-Compton opacity:
$\sigma_{\ApJstyle\rm C}^{\ApJstyle\rm s}({\rm iso})
 /\sigma_{\ApJstyle\rm C}({\rm iso})
=
\left[\sigma_{\ApJstyle\rm C}^{\ApJstyle\rm s}-
\sigma_{\ApJstyle\rm C}^{\ApJstyle\rm s}
  \left(\theta_{\ApJstyle\rm f}\right)\right]
/\left[\sigma_{\ApJstyle\rm C}-
  \sigma_{\ApJstyle\rm C}^{\ApJstyle\rm s}
  \left(\theta_{\ApJstyle\rm f}\right)\right]$.

     There seems no precise way of optimizing the iso-Compton
opacity:  the smaller the cone's opening angle, the more
the forward component represents truly forward and coherent
scattering, but
the less one exploits the forward peaking of Compton opacity.
We suggest the following prescription for
$\sigma_{\ApJstyle\rm C}^{\ApJstyle\rm s}
  \left(\theta_{\ApJstyle\rm f}\right)$
which allows us to explore the options: 
$$\sigma_{\ApJstyle\rm C}^{\ApJstyle\rm s}
    \left(\theta_{\ApJstyle\rm f}\right)
 =\left(\sigma_{\ApJstyle\rm C,f}^{\ApJstyle\rm s}
        -\sigma_{\ApJstyle\rm C,b}^{\ApJstyle\rm s}\right)
  \min\left(g,1\right)
  +2\sigma_{\ApJstyle\rm C,b}^{\ApJstyle\rm s}\max\left(g-1,0\right)
                                                  \,\, ,\eqno(\eqn)$$
where 
$\sigma_{\ApJstyle\rm C,f}^{\ApJstyle\rm s}$
and
$\sigma_{\ApJstyle\rm C,b}^{\ApJstyle\rm s}$ are the partial
Compton scattering cross sections for the front and
back scattering hemispheres, respectively,
and $g\in[0,2]$ is an adjustable parameter.
The values for 
$\sigma_{\ApJstyle\rm C,f}^{\ApJstyle\rm s}$
and
$\sigma_{\ApJstyle\rm C,b}^{\ApJstyle\rm s}$
can be obtained from 
$$\eqalignno{
 \sigma_{\ApJstyle\rm C}^{\ApJstyle\rm s}(\theta)
   &=\sigma_{\ApJstyle e}\left({3\over8}\right)
    \Biggl\{
     {\ln\left(1+\alpha-\alpha\cos\theta\right)
       \over\alpha^{\ApJstyle 3}}                       \cr
   &\qquad\qquad\qquad -\left[6\alpha^{\ApJstyle 2}
                  \left(1+\alpha-\alpha\cos\theta\right)^{\ApJstyle 3}
                    \right]^{\ApJstyle -1} \cr
   &\qquad\qquad\qquad\qquad
\biggl[6+15\alpha+3\alpha^{\ApJstyle 2}-12\alpha^{\ApJstyle 3}
                                       -8\alpha^{\ApJstyle 4} \cr
   &\qquad\qquad\qquad\qquad\quad
      -\left(6+30\alpha+27\alpha^{\ApJstyle 2}-18\alpha^{\ApJstyle 3}
                          -24\alpha^{\ApJstyle 4}\right)\cos\theta \cr
   &\qquad\qquad\qquad\qquad\quad
      +\left(15\alpha+33\alpha^{\ApJstyle 2}
           -24\alpha^{\ApJstyle 4}\right)\cos^{\ApJstyle 2}\theta \cr
   &\qquad\qquad\qquad\qquad\quad
      -\left(9\alpha^{\ApJstyle 2}+6\alpha^{\ApJstyle 3}
    -8\alpha^{\ApJstyle 4}\right)\cos^{\ApJstyle 3}\theta \biggr]\Biggr\}
                               &(\eqn) \cr}$$
(e.g., Davisson 1965, p.~55) which
is the Compton scattering opacity for the scattering cone of
opening angle $\theta$ centered on the forward direction:
$\sigma_{\ApJstyle\rm C,f}^{\ApJstyle\rm s}=
\sigma_{\ApJstyle\rm C}^{\ApJstyle\rm s}\left(\pi/2\right)$
and
$\sigma_{\ApJstyle\rm C,b}^{\ApJstyle\rm s}=
\sigma_{\ApJstyle\rm C}^{\ApJstyle\rm s}-
\sigma_{\ApJstyle\rm C}^{\ApJstyle\rm s}\left(\pi/2\right)$.
Note that
$\sigma_{\ApJstyle\rm C,f}^{\ApJstyle\rm s}-
\sigma_{\ApJstyle\rm C,b}^{\ApJstyle\rm s}$
rises from 0 at $\alpha=0$ to a
maximum of $\sim 0.19479908\sigma_{\ApJstyle e}$
at $\alpha=0.49083380$ and
then declines to 0 at $\alpha=\infty$.

     The point of equation~(\showeqn -1) is that the special
$g$ values 0, 1, and 2 give
$\sigma_{\ApJstyle\rm C}^{\ApJstyle\rm s}
    \left(\theta_{\ApJstyle\rm f}\right)$
expressions with identifiable physical significance.
If $g=0$, then
$\sigma_{\ApJstyle\rm C}^{\ApJstyle\rm s}
    \left(\theta_{\ApJstyle\rm f}\right)=0$
and we have the angle-averaged Compton opacity approximation that
we describe below.
If $g=1$, then
$\sigma_{\ApJstyle\rm C}^{\ApJstyle\rm s}
    \left(\theta_{\ApJstyle\rm f}\right)=
\sigma_{\ApJstyle\rm C,f}^{\ApJstyle\rm s}
        -\sigma_{\ApJstyle\rm C,b}^{\ApJstyle\rm s}$
and the iso-Compton opacity will equal what can plausibly be 
identified as the real isotropic component of Compton opacity.
If $g=2$, then
$\sigma_{\ApJstyle\rm C}^{\ApJstyle\rm s}
    \left(\theta_{\ApJstyle\rm f}\right)=
\sigma_{\ApJstyle\rm C,f}^{\ApJstyle\rm s}
        +\sigma_{\ApJstyle\rm C,b}^{\ApJstyle\rm s}
=\sigma_{\ApJstyle\rm C}^{\ApJstyle\rm s}$
and the iso-Compton opacity approximation reduces to the
absorption-only approximation that we discussed above.
We note that when a $\gamma$-ray's energy is zero,
equation~(\showeqn -1) yields a forward component
of zero for all $g\leq1$ since
$\sigma_{\ApJstyle\rm C,f}^{\ApJstyle\rm s}
        =\sigma_{\ApJstyle\rm C,b}^{\ApJstyle\rm s}$
for $\alpha=0$.

     The quantity $g$ could be considered a free parameter.
Given a highly accurate $\gamma$-ray deposition calculation for
comparison, one could perhaps fine tune $g$ to make the
LS~procedure yield a highly accurate result.
But we are seeking a procedure without free parameters.
Thus we have chosen to use hereafter (except where we
explicitly say otherwise) the $g$ value that
{\it a~priori} seems most reasonable:  i.e.,
$$ g=1 
\qquad{\rm giving}\qquad
\sigma_{\ApJstyle\rm C}^{\ApJstyle\rm s}
    \left(\theta_{\ApJstyle\rm f}\right)=
\sigma_{\ApJstyle\rm C,f}^{\ApJstyle\rm s}-
\sigma_{\ApJstyle\rm C,b}^{\ApJstyle\rm s}
\,\, .\eqno(\eqn)$$
Some numerical experimentation suggests that $g=1$ 
will quite accurately reproduce the results of the optimized
SSH~procedure (see \S~6) and that $g$ widely different from
1 will not.

     One can equate the 
$\sigma_{\ApJstyle\rm C}^{\ApJstyle\rm s}(\theta_{\ApJstyle\rm f})$
(from eq.~[\showeqnsq -0]))
to the right-hand side of equation~(\showeqn -1)
and solve numerically (e.g., by a Newton-Raphson iteration)
for $\theta_{\ApJstyle\rm f}$.
The $\theta_{\ApJstyle\rm f}$ value
increases monotonically
with energy:  it is $0^{\circ}$ for $\alpha=0$ and goes to an
asymptotic value of $\arccos(1/3)\approx70.528779^{\circ}$ for
$\alpha=\infty$.
With this $\theta_{\ApJstyle\rm f}$, one can compute
the energy reduction factors
$\sigma_{\ApJstyle\rm C}^{\ApJstyle\rm s}
  \left(\theta_{\ApJstyle\rm f}\right)
/\sigma_{\ApJstyle\rm C}
  \left(\theta_{\ApJstyle\rm f}\right)$
and
$\left[\sigma_{\ApJstyle\rm C}^{\ApJstyle\rm s}-
\sigma_{\ApJstyle\rm C}^{\ApJstyle\rm s}
  \left(\theta_{\ApJstyle\rm f}\right)\right]
/\left[\sigma_{\ApJstyle\rm C}-
  \sigma_{\ApJstyle\rm C}
  \left(\theta_{\ApJstyle\rm f}\right)\right]$.
One can then compare $\theta_{\ApJstyle\rm f}$ and these
energy reduction factors to $0^{\circ}$, 1, and
$\sigma_{\ApJstyle\rm C}^{\ApJstyle\rm s}({\rm iso})
 /\sigma_{\ApJstyle\rm C}({\rm iso})$.
The closer the agreement, the better the iso-Compton opacity
approximation represents the actual Compton opacity.
The agreement is best for $\alpha=0$ and degrades
monotonically as $\alpha$ increases.
For $\alpha=2$ ($\approx1\,$MeV),
$\theta_{\ApJstyle\rm f}\approx63.01^{\circ}$,
$\sigma_{\ApJstyle\rm C}^{\ApJstyle\rm s}
  \left(\theta_{\ApJstyle\rm f}\right)
/\sigma_{\ApJstyle\rm C}
  \left(\theta_{\ApJstyle\rm f}\right)
\approx0.7457$,
and
$\left[\sigma_{\ApJstyle\rm C}-\sigma_{\ApJstyle\rm C}
      \left(\theta_{\ApJstyle\rm f}\right)\right]
  /\sigma_{\ApJstyle\rm C}({\rm iso})\approx0.7537$.
For $\alpha=8$ ($\approx4\,$MeV), the
corresponding values are
$68.07^{\circ}$, $0.5243$, and $0.4929$.
By these criteria the iso-Compton opacity approximation
can only be expected to be moderately successful for
the unscattered decay $\gamma$-rays in supernovae, but that it should
be better for lower energy, scattered $\gamma$-rays.
The fact that the Compton  
absorption
component increases with $\gamma$-ray energy
(e.g., its fractional value is
 0.4431 for $\alpha=2$ and
 0.6091 for $\alpha=8$), however, limits the error
in treating the scattering using the iso-Compton
opacity approximation as $\gamma$-ray energy increases. 

    As mentioned above, in the limit of $\alpha$ going to 0,
Compton scattering reduces to coherent Thomson scattering.
Thomson scattering, albeit not isotropic, is symmetric about the
plane perpendicular to the forward direction (e.g., Mihalas 1978, p.~30),
and thus
$\sigma_{\ApJstyle\rm C}^{\ApJstyle\rm s}
 \left(\theta_{\ApJstyle\rm f}\right)$
(for $g=1$ of course)
goes to 0 when $\alpha$ goes to 0.
Therefore, iso-Compton opacity
also reduces to coherent scattering and
its value becomes equal to that of Compton opacity when $\alpha$
goes to 0.
But iso-Compton scattering does not actually reduce to Thomson scattering
since it remains isotropic.
Thomson scattering, however, in most practical calculations can be
approximated as isotropic scattering.
Therefore, iso-Compton scattering reduces to a usual good approximation
to Thomson scattering in the limit of $\alpha$ going to 0.

     A simpler alternative to the iso-Compton opacity approximation
is the angle-averaged Compton opacity approximation
(i.e., the iso-Compton opacity approximation for $g=0$).
In this latter approximation, one just assumes Compton opacity
acts like its angle average:  i.e., the scattering is isotropic and
the energy reduction factor on scattering,
$\sigma_{\ApJstyle\rm C}^{\ApJstyle\rm s}/\sigma_{\ApJstyle\rm C}$,
is a constant with angle.
Since the angle-averaged Compton opacity approximation does not exploit
the forward peaking of Compton opacity, it is a poorer approximation
from our point of view than the iso-Compton opacity approximation.
In principle, it is clear that a two-component approximation to
the angle-dependence of Compton opacity, if well chosen, should be
a better approximation than a one-component approximation.
\vskip 2\baselineskip

\centerline{3.\ \  THE MEAN OPACITIES}\nobreak
\vskip\baselineskip\nobreak
     A grey atmosphere is an atmosphere with a frequency-independent
opacity (e.g., Mihalas 1978, p.~53ff).
In a grey atmosphere, only the frequency-integrated radiation
field is needed to solve the radiative transfer problem:
this is a great simplification.
One can imagine finding a mean opacity that will reduce a non-grey
problem to a grey one.
For practical purposes, however, (with the important exception of
radiative transfer in the diffusion limit with local thermodynamic
equilibrium [LTE])
no mean opacity permits a non-grey problem to be completely
reduced to a grey problem (e.g., Mihalas 1978, p.~56ff).
This is mainly because the mean opacity that would perform the
reduction exactly cannot usually be calculated until after the problem
is solved.
If one wants to use grey radiative transfer as an approximation, one
tries to find the mean opacity that {\it a~priori} offers the best chance
for an accurate solution for those effects that are of particular
interest.
This is the kind of mean opacity we try to find in this section.
\vskip\baselineskip

\centerline{3.1\ \  \it Mean Opacity Prescriptions}\nobreak  
\vskip .5\baselineskip\nobreak
     What we want ultimately is to find the $\gamma$-ray
energy deposition.
Thus, we want to get as accurately as possible the amount of
energy absorbed from the $\gamma$-ray fields.
These $\gamma$-ray fields arise directly from the radioactive decay
and from scattering.
To get the scattered $\gamma$-ray fields we need the emissivity
provided by scattering and this is obtained from the amount of
flux from a beam removed by scattering (see \S~4).
Therefore, we need to consider scattering opacity as well as absorption
opacity.

     To find the appropriate opacities consider the flux removed at
a point either by absorption, scattering, or both processes.
For clarity, quantities not at the removal point (non-local quantities) will
be distinguished by a functional dependence on the beam path length 
$s$ which is zero at the removal point.
Quantities at the removal point will not be given an explicit
position dependence.
Thus $h(s)$ is the $h$-quantity at location $s$ and not at the removal
point, and $h=h(0)$ is the $h$-quantity at the removal point.
Making the approximation of time-independent, static radiative
transfer,
the energy removed at the removal point at a given frequency
$\nu$ from a beam of specific intensity $I_{\ApJstyle\nu}$ is given by
$$ \chi_{\ApJstyle\nu}^{\ApJstyle R}I_{\ApJstyle\nu}=
\chi_{\ApJstyle\nu}^{\ApJstyle R}
I_{\ApJstyle\nu}\left(s_{\ApJstyle\rm b}\right)
    \exp\left[-\tau_{\ApJstyle\nu}\left(s_{\ApJstyle\rm b}\right)\right]
+\chi_{\ApJstyle\nu}^{\ApJstyle R}
 \int_{\ApJstyle 0}^{\ApJstyle s_{\ApJstyle\rm b}}ds\,
   \eta_{\ApJstyle\nu}\left(s\right)
    \exp\left[-\tau_{\ApJstyle\nu}\left(s\right)\right] 
                                                    \,\, , \eqno(\eqn)$$
where $\chi_{\ApJstyle\nu}^{\ApJstyle R}$ is the
frequency-specific extinction at the removal point and $R$ (as
specified in \S~2) is blank, ``a'', and ``s'' for
total, absorption, and scattering extinction, respectively.
The location of a boundary on the beam path is given by
$s_{\ApJstyle\rm b}$.
The $I_{\ApJstyle\nu}\left(s_{\ApJstyle\rm b}\right)$ is the specific
intensity incident on the boundary,
$\tau_{\ApJstyle\nu}\left(s_{\ApJstyle\rm b}\right)$ is the frequency-specific
optical depth to the boundary,
$\eta_{\ApJstyle\nu}\left(s\right)$ is the 
frequency-specific emissivity along the beam path,
and 
$\tau_{\ApJstyle\nu}\left(s\right)$ is the frequency-specific
optical depth to $s$.
Note that the optical depth is calculated using total extinction.
Also note that
$\chi_{\ApJstyle\nu}^{\ApJstyle R}=\rho\kappa_{\ApJstyle\nu}^{\ApJstyle R}$
where
$\rho$ is the density at the removal point and
$\kappa_{\ApJstyle\nu}^{\ApJstyle R}$ is
the frequency-specific opacity at the removal point.
To get the total energy removed by either absorption, scattering, or both
processes), we integrate equation~(\showeqn -0)
over all frequency and obtain
$$ \eqalignno{
\int_{\ApJstyle 0}^{\ApJstyle\infty}d\nu\,
\chi_{\ApJstyle\nu}^{\ApJstyle R}I_{\ApJstyle\nu}&=
\int_{\ApJstyle 0}^{\ApJstyle\infty}d\nu\,\chi_{\ApJstyle\nu}^{\ApJstyle R}
I_{\ApJstyle\nu}\left(s_{\ApJstyle\rm b}\right)
  \exp\left[-\tau_{\ApJstyle\nu}\left(s_{\ApJstyle\rm b}\right)\right] \cr
&\qquad
+\int_{\ApJstyle 0}^{\ApJstyle\infty}d\nu\,\chi_{\ApJstyle\nu}^{\ApJstyle R}
 \int_{\ApJstyle 0}^{\ApJstyle s_{\ApJstyle\rm b}}ds\,
   \eta_{\ApJstyle\nu}\left(s\right)
    \exp\left[-\tau_{\ApJstyle\nu}\left(s_{\ApJstyle\rm b}\right)\right]   
                                                          \,\, . &(\eqn) \cr}$$
The grey equation, with which we want to replace equation~(\showeqn -0),
is
$$ \chi^{\ApJstyle R} I=
\chi^{\ApJstyle R} I\left(s_{\ApJstyle\rm b}\right)
  \exp\left[-\tau\left(s_{\ApJstyle\rm b}\right)\right]
+\chi^{\ApJstyle R}
  \int_{\ApJstyle 0}^{\ApJstyle s_{\ApJstyle\rm b}}ds\, \eta\left(s\right)
                \exp\left[-\tau\left(s\right)\right]          \,\, , \eqno(\eqn)$$
where
$$\eqalignno{
 I&=\int_{\ApJstyle 0}^{\ApJstyle\infty}d\nu\, I_{\ApJstyle\nu} \,\, , &(\eqn) \cr
 I\left(s_{\ApJstyle\rm b}\right)&=\int_{\ApJstyle 0}^{\ApJstyle\infty}d\nu\,
               I_{\ApJstyle\nu}\left(s_{\ApJstyle\rm b}\right) \,\, , &(\eqn) \cr
\noalign{\noindent\rm and}
 \eta(s)&=\int_{\ApJstyle 0}^{\ApJstyle\infty}d\nu\,
             \eta_{\ApJstyle\nu}\left(s\right)
                                                            \,\, . &(\eqn) \cr}$$
The $\chi^{\ApJstyle R}$ is the mean extinction at the removal point
and $\tau(s)$ is the mean optical depth back along the beam path.

     Just by equating the left-hand sides of
equations~(\showeqn -4) and~(\showeqn -3) and trying to solve for
$\chi^{\ApJstyle R}$ it becomes clear that the mean opacity that will render
equations~(\showeqn -4) and~(\showeqn -3) exactly equivalent
will have a complex dependence on non-local quantities through
$I_{\ApJstyle\nu}$.
Such a mean opacity would be useless practically since we would have
to do the frequency-specific radiative transfer to evaluate it.
To obtain a practical mean opacity, we need to make some approximations.

     First, let us try to find a mean opacity that will force
the agreement of the first
terms on the right-hand sides of equations~(\showeqn -4) and~(\showeqn -3).
Such an opacity would be obtained from
$$ \kappa^{\ApJstyle R}\exp\left[-\tau\left(s_{\ApJstyle\rm b}\right)\right]=
{ \int_{\ApJstyle 0}^{\ApJstyle\infty}d\nu\,
  I_{\ApJstyle\nu}\left(s_{\ApJstyle\rm b}\right)
   \kappa_{\ApJstyle\nu}^{\ApJstyle R}
   \exp\left[-\tau_{\ApJstyle\nu}\left(s_{\ApJstyle\rm b}\right)\right]
 \over 
 \int_{\ApJstyle 0}^{\ApJstyle\infty}d\nu\,
  I_{\ApJstyle\nu}\left(s_{\ApJstyle\rm b}\right)
}                                            \,\, .    \eqno(\eqn)$$
Equation~(\showeqn -0) is clearly a highly non-linear, non-local equation
for determining $\kappa^{\ApJstyle R}$ at all points in the atmosphere.
To make progress let us assume the optical depth is very small at all
frequencies and then the exponential factors can be set to 1.
Let us further assume that beyond the boundary, the incident
radiation field forms in an optically thin layer of spatial width
$\Delta s$ and spatially constant emissivity.
This assumption gives
$I_{\ApJstyle\nu}\left(s_{\ApJstyle\rm b}\right)
  =\eta_{\ApJstyle\nu}\left(s_{\ApJstyle\rm b}\right)\Delta s$.
Equation~(\showeqn 0) now reduces to
$$ \kappa^{\ApJstyle R}=
{ \int_{\ApJstyle 0}^{\ApJstyle\infty}d\nu\,
  \eta_{\ApJstyle\nu}\left(s_{\ApJstyle\rm b}\right)
  \kappa_{\ApJstyle\nu}^{\ApJstyle R}
 \over 
 \int_{\ApJstyle 0}^{\ApJstyle\infty}d\nu\,
  \eta_{\ApJstyle\nu}\left(s_{\ApJstyle\rm b}\right)
}                                            \,\, .    \eqno(\eqn)$$
Equation~(\showeqn 0) is formally a non-local definition of the mean
opacity since
the emissivities are to be evaluated on the boundary not at the
removal point.
In the supernova $\gamma$-ray case, however, the emissivity of the
{unscattered} $\gamma$-ray field  
(which is provided by radioactive species)
is constant in space 
aside from the frequency-independent factor of radioactive
species number density (see eq.~[\showeqnsq +4] below).
We will assume that the emissivity of a 
{scattered} $\gamma$-ray   
field can also be approximated constant in space 
aside from frequency-independent scale factors.
From the above considerations, equation~(\showeqn -0) can be rewritten
$$ \kappa^{\ApJstyle R}=
{ \int_{\ApJstyle 0}^{\ApJstyle\infty}d\nu\,
  \eta_{\ApJstyle\nu}
  \kappa_{\ApJstyle\nu}^{\ApJstyle R}
 \over 
 \int_{\ApJstyle 0}^{\ApJstyle\infty}d\nu\,
  \eta_{\ApJstyle\nu}
}                                            \,\, .    \eqno(\eqn)$$
We will call equation~(\showeqn -0) the emissivity-weighted mean
opacity.
From our derivation, the emissivity-weighted mean opacity
is strictly valid only in an optically thin medium where
the incident beams are formed in an optically thin, non-local emission
region and only for an emissivity whose frequency behavior is
constant in space.

     Let now us try to find a mean opacity that will force the agreement
of the second 
terms on the right-hand sides of equations~(\showeqn -7) and~(\showeqn -6).
We first make the assumption that the density, emissivities, and
opacities are
nonzero constants with respect to location in a region of space that encloses
the removal point and are zero outside of this region.
This means, of course, that the removal point is embedded in a region
of $\gamma$-ray emission. 
Then from the second terms we obtain the following expression:
$$ {\kappa^{\ApJstyle R}\over\kappa}
   \left\{1-\exp\left[-\tau\left(s_{\ApJstyle\rm b}\right)\right]\right\}
={\int_{\ApJstyle 0}^{\ApJstyle\infty}d\nu\,
  \eta_{\ApJstyle\nu}
  {\displaystyle {\kappa_{\ApJstyle\nu}^{\ApJstyle R}
    \over\kappa_{\ApJstyle\nu}} }
   \left\{1-\exp\left[-\tau_{\ApJstyle\nu}\left(s_{\ApJstyle\rm b}\right)
      \right]\right\}
  \over
   \int_{\ApJstyle 0}^{\ApJstyle\infty}d\nu\,\eta_{\ApJstyle\nu}
 }                                                     \,\, . \eqno(\eqn)$$
Note that we had to introduce the total extinctions $\chi$ and
$\chi_{\ApJstyle\nu}$ in order to form the differentials
$d\tau=ds\,\chi$ and $d\tau_{\ApJstyle\nu}=ds\,\chi_{\ApJstyle\nu}$
for the optical depth integrations.
This is where the ${\kappa^{\ApJstyle R}/\kappa}$ and 
${\kappa_{\ApJstyle\nu}^{\ApJstyle R}/\kappa_{\ApJstyle\nu}}$
ratios come from.
These ratios, of course, become 1 for the total opacity case:
i.e., when $R$ is blank.

     Equation~(\showeqn -0) is almost a local expression for mean
opacity since the only non-local dependence is on the distance 
to the boundaries. 
It is, however, highly nonlinear.
But certain things can be said about it.
First, in the optically thin limit, which is an important supernova case,
equation~(\showeqn -0) reduces to equation~(\showeqn -1).
Second, in intermediate optically thick cases where the range
in variation of the optical depth with frequency is $\lapprox 1$, 
equation~(\showeqn -0) reduces approximately to equation~(\showeqn -1).
This situation can roughly arise for the unscattered $\gamma$-ray
fields of $^{\ApJstyle 56}$Co and $^{\ApJstyle 56}$Ni
(see SSH's Fig.~1).
Third, in the optically thick limit in supernovae, all $\gamma$-rays
are locally trapped and eventually absorbed (see \S~4), and so
the exact values of the mean opacity and its components
are not important for deposition calculations provided only that
they are sufficiently large.
In this case, equation~(\showeqn -0) can be replaced adequately
by any reasonable mean opacity prescription.
Given these facts, the emissivity-weighted mean opacity
equation~(\showeqn -1) seems a plausible and practical replacement
for equation~(\showeqn -0).


     It is true that there are cases where the emissivity-weighted mean
opacity is a poor choice.
For instance, consider equation~(\showeqn -3) for the mean opacity when the
removal point is outside of the $\gamma$-ray emission region.
Assume that the optical depths to the emission region
(i.e., to the boundary)
are small, but that the emission region is optically thick.
In this case, from simple radiative transfer
$I_{\ApJstyle\nu}\left(s_{\ApJstyle\rm b}\right)=
\eta_{\ApJstyle\nu}\left(s_{\ApJstyle\rm b}\right)
 /\chi_{\ApJstyle\nu}\left(s_{\ApJstyle\rm b}\right)$
assuming the emission region has spatially constant
emissivity and extinction.
If we now assume that the opacities (but not necessarily density)
are constant in
space and assume that emissivities are spatially constant aside from
frequency-independent factors, then we obtain from equation~(\showeqn -3)
$$ \kappa^{\ApJstyle R}=
{ \int_{\ApJstyle 0}^{\ApJstyle\infty}d\nu\,
  {\displaystyle 
  \eta_{\ApJstyle\nu} 
  {\kappa_{\ApJstyle\nu}^{\ApJstyle R} \over \kappa_{\ApJstyle\nu}}
  }
 \over
 \int_{\ApJstyle 0}^{\ApJstyle\infty}d\nu\,
  {\displaystyle \eta_{\ApJstyle\nu}{1\over\kappa_{\ApJstyle\nu}} }
}                                            \,\, .    \eqno(\eqn)$$
In this special case, we find that emissivity-weighted 
inverse-mean opacities are the best choice for the mean opacities.
The values of these mean opacities tend to be dominated by the
lowest frequency-specific total opacities.
The reason is that lowest total opacities, permit greatest transfer of flux
in the optically thick emission region.
The emissivity-weighted inverse-mean opacities may not be too important in
supernovae.
They apply in optically thin regions when the emission is occurring
in optically thick regions.
But in such cases, most $\gamma$-ray deposition in the supernova tends
to occurs in the optically thick regions and this energy flows out
into the optically thin regions in the form of UV-optical-IR
radiation which may mainly determine the thermal, ionization, and
excitation energy in those optically thin regions.
The emissivity weighted inverse-mean opacity prescription is, of course,
only one of many special prescriptions that can be invented for special
cases.

      The preceding analysis has led us to choose the emissivity-weighted
mean opacity prescription given by equation~(\showeqn -2) as
the mean opacity prescription for the LS~procedure.
The emissivity-weighted mean opacities become exactly right for
supernovae in the optically thin limit (when scattered fields are not
important)
whether the removal point is embedded in a region of emission or not.
In the optically thick limit when the removal point is embedded
in an emission region, they capture some
of the right behavior and can do no harm if all the $\gamma$-rays
are locally trapped and absorbed.
The cases where they are a poor approximation may not be so
important.
The emissivity-weighted mean opacities are also straightforward to
calculate.

     A factor that may further limit the error in using the
emissivity-weighted mean opacities in cases for which they are 
formally not optimum is
that the dominant $\gamma$-ray opacity in supernovae does not
vary strongly with frequency across the frequency band where
it is most important.
Going from $0.1\,$MeV to $4\,$MeV,
Compton total opacity falls by only a factor of $\sim 5$
and iso-Compton total opacity by only a factor of $\sim 6$.
Since $\gamma$-ray spectra from radioactive decays are
often dominated by one or a few $\gamma$-ray lines,
it is possible that almost all the emissivity weighting will be
given to frequency-specific opacities that are rather close
in value since they come from a rather narrow frequency band.
If this is so, then the difference between the emissivity-weighted
mean opacities, emissivity-weighted inverse-mean opacities,
or any other kind of reasonable mean opacity with emissivity weighting
may often be rather small for the unscattered $\gamma$-ray field.
This expectation is, in fact, fulfilled insofar as we have
tested it (see \S~3.2).
The difference between the emissivity-weighted
mean and emissivity-weighted inverse-mean opacities turns out
to be even smaller for the scattered $\gamma$-ray fields when
iso-Compton opacity is used as we will show in \S~3.2.
\vskip\baselineskip

\centerline{3.2\ \  \it Mean Opacities for the LS Procedure}\nobreak
\vskip .5\baselineskip\nobreak
     For the LS~procedure we will need emissivity-weighted
mean opacities (which we will usually just call mean opacities
hereafter) and
some other mean quantities for multiple orders of scattered
$\gamma$-ray fields.
The 0th order field is the field emergent from the radioactive
decay itself.
The higher order fields, 1st, 2nd, 3rd, etc., have undergone 1, 2, 3, etc., 
scattering events.

     To obtain the mean quantities we will make a number of
sweeping assumptions.
The general rationale for proceeding despite the deficiencies
in these and the other assumptions we have or will make
is that our assumptions allow the LS~procedure formalism
we develop to capture some of the correct physical behavior
while remaining fairly simple.
The accuracy of the LS~procedure must be verified by comparison
to procedures of known accuracy.
A limited comparison of this sort is done in \S~6.

     The most sweeping assumption we will make is that
for the purposes of the derivations the medium can be
considered as infinite (which implies that the medium is
in the optically thick limit), homogeneous, isotropic, and
time-independent.
This implies that we also consider the $\gamma$-ray fields
to be homogeneous, isotropic, and time-independent.
We emphasize that we will apply the mean opacities we derive
from the assumed state of the medium to cases where the medium
is not in the assumed state and that the assumed state
is not the optimum state for the application of the
emissivity-weighted mean opacity prescription (see \S~3.1).
The aforementioned general rationale allows us to proceed
anyway.

     The radioactive decay $\gamma$-ray spectrum is virtually
entirely a line spectrum and we will assume it is exactly
so.
Because of our iso-Compton opacity and pair
production opacity approximations, the higher order fields
will also then consist of line spectra in our treatment.
Because we are dealing only with line spectra,
we will from now on use line-integrated or line-mean
quantities and do sums over these quantities (instead of
integrals) to obtain the mean
quantities we need for the LS~procedure. 

     The 0th order $\gamma$-ray emissivity of a line $j$ from
some radioactive species is given by
$$  \eta_{\ApJstyle 0,j}^{\ApJstyle\gamma}
   ={1\over 4\pi}{n\over t_{\ApJstyle e}}
     f_{\ApJstyle0,j}E_{\ApJstyle 0,j}    \,\, , \eqno(\eqn)$$
where $n$ is the number density of the species,
$t_{\ApJstyle e}$ is the $e$-folding time for the decay, 
$f_{\ApJstyle0,j}$ is the number of $\gamma$-rays (which is usually
less than 1) in line $j$ per decay,
and $E_{\ApJstyle0,j}$ is the $\gamma$-ray line energy.
In order to obtain the 0th order mean opacities we in fact need
only the $E_{\ApJstyle 0,j}$ and $f_{\ApJstyle0,j}$ quantities
since the other emissivity factors cancel out
(see eq.~[\showeqnsq -3]).

      Based on our assumption of line spectra in all scattering
orders, we will have definite
$E_{\ApJstyle i,j}$'s and $f_{\ApJstyle i,j}$'s for
orders $i>0$.
Moreover, the frequency dependence of the emissivity in all
orders will be given by $f_{\ApJstyle i,j}E_{\ApJstyle i,j}$ as we
will show below.
Thus, we can use the
$E_{\ApJstyle i,j}$'s and $f_{\ApJstyle i,j}$'s
to calculate the mean opacities for all orders.
To obtain $E_{\ApJstyle i,j}$'s and $f_{\ApJstyle i,j}$'s
we will derive recurrence relations.
In order to have simple recurrence relations we need to keep
constant the number of lines through all scatterings.
Therefore we assume that we can replace the three opacities
(pair production, photoelectric, and iso-Compton) by a single
combined opacity.
(Note the combined opacity is a sum of the opacity types with
mean properties, not a frequency mean of opacity.)
First let us obtain the recurrence relation for the
$E_{\ApJstyle i,j}$'s.

     The energy of a scattered $\gamma$-ray (assuming there
is no angle-dependence on the scattered energy) is usually
$\kappa^{\ApJstyle\rm s}/\kappa$.
This applies to the iso-Compton opacity.
In a case like that of the pair production opacity where
we have assumed that two scattered $\gamma$-rays share equally
the energy of one incident $\gamma$-ray and where there is no
angle-dependence for the scattered energy,
the energy of each scattered $\gamma$-ray is
$(1/2)\kappa^{\ApJstyle\rm s}/\kappa$.
Now for our combined opacity the number (not energy)
fractions of scattered $\gamma$-rays that are scattered by
the iso-Compton and pair production opacities 
are 
$$ {\kappa({\rm iso}) \over
    \kappa({\rm iso})+\kappa({\rm pair}) }
   \qquad{\rm and}\qquad 
    {\kappa({\rm pair}) \over
    \kappa({\rm iso})+\kappa({\rm pair}) }
                          \,\, ,                \eqno(\eqn)$$
respectively,
where ``${\rm iso}$'' stands for the iso-Compton opacity
and ``pair'' for pair production opacity.
Therefore the recurrence relation giving the mean energy of
$\gamma$-rays in line~$j$ for all scattered orders is
$$
\eqalignno{
E_{\ApJstyle i,j}
&=E_{\ApJstyle i-1,j}
\Bigggl\{
 {\kappa_{\ApJstyle i-1,j}\left({\rm iso}\right)
 \over
   \kappa_{\ApJstyle i-1,j}\left({\rm iso}\right)
  +\kappa_{\ApJstyle i-1,j}\left({\rm pair}\right) }
   \left[
   {\kappa_{\ApJstyle i-1,j}^{\ApJstyle\rm s}\left({\rm iso}\right)
    \over
    \kappa_{\ApJstyle i-1,j}\left({\rm iso}\right)}
    \right] \cr
&\qquad\qquad\qquad +
 {\kappa_{\ApJstyle i-1,j}\left({\rm pair}\right)
 \over
   \kappa_{\ApJstyle i-1,j}\left({\rm iso}\right)
  +\kappa_{\ApJstyle i-1,j}\left({\rm pair}\right) }
   \left[{1\over2}
   {\kappa_{\ApJstyle i-1,j}^{\ApJstyle\rm s}\left({\rm pair}\right)
    \over
    \kappa_{\ApJstyle i-1,j}\left({\rm pair}\right)}
   \right]
\Bigggr\}  \cr
&=E_{\ApJstyle i-1,j}
\left[
{ \displaystyle
  \kappa_{\ApJstyle i-1,j}^{\ApJstyle\rm s}\left({\rm iso}\right)
 +\left({1\over2}\right)
    \kappa_{\ApJstyle i-1,j}^{\ApJstyle\rm s}\left({\rm pair}\right)
 \over
   \kappa_{\ApJstyle i-1,j}\left({\rm iso}\right)
  +\kappa_{\ApJstyle i-1,j}\left({\rm pair}\right)
}
\right]  \,\, .                              &(\eqn) \cr}$$
Note that the photoelectric opacity is assumed to be a pure
absorption opacity, and therefore has no effect on the energy of
scattered $\gamma$-rays.

     Now we need to determine the $f_{\ApJstyle i,j}$'s.
Consider a single decay and imagine tracking decay-emitted
$\gamma$-rays through all scattering events.
Because the medium is assumed to be infinite, homogeneous,
and isotropic we can treat all the $\gamma$-rays in line~$j$
from the single decay together as a single packet.
Because the medium is infinite each packet emitted in any
order will eventually be scattered:  the distance and travel
time will depend on the extinction in the order.
In other words, if $f_{\ApJstyle i,j}$ $\gamma$-rays are
emitted into an order, then $f_{\ApJstyle i,j}$ $\gamma$-rays
will be removed from the order, either scattered or absorbed.
Consider the packet with $\gamma$-ray number $f_{\ApJstyle i-1,j}$.
On the $(i-1)$th scattering event with the combined opacity
the packet's $\gamma$-ray number will change according to 
$$f_{\ApJstyle i,j}=f_{\ApJstyle i-1,j}
\left[
{ \displaystyle
   \kappa_{\ApJstyle i-1,j}\left({\rm iso}\right)
 +2\kappa_{\ApJstyle i-1,j}\left({\rm pair}\right)
 \over
   \kappa_{\ApJstyle i-1,j}\left({\rm iso}\right)
  +\kappa_{\ApJstyle i-1,j}\left({\rm pair}\right)
  +\kappa_{\ApJstyle i-1,j}\left({\rm pe}\right)
}
\right]  \,\, ,                                    \eqno(\eqn)$$
where ``pe'' stands for photoelectric opacity and
where the denominator in the fraction accounts for the $\gamma$-rays
removed from the $(i-1)$th order field and the numerator for the
$\gamma$-rays emitted into the $i$th field.
Equation~(\showeqn -0) is the recurrence relation for
the $f_{\ApJstyle i,j}$'s.
Now consider all decays in the medium.
Because the medium is infinite, homogeneous, isotropic, and
time-independent,
the time and distance a packet took in going from the decay
emission to $(i-1)$th order scattering event is irrelevant to
the emissivity in the $i$th order.
Thus, the emissivity for line~$j$ in the $i$ order will be
proportional to
$f_{\ApJstyle i,j}E_{\ApJstyle i,j}$ and will not have any
other frequency dependence, and will be constant in space.

     Using equation~(\showeqn -6) with the integrals converted to sums
and given the $E_{\ApJstyle i,j}$ and $f_{\ApJstyle i,j}$ values,
the expression for $i$th order emissivity-weighted mean opacities is
found to be
$$  \kappa_{\ApJstyle i}^{\ApJstyle R}=
{\sum_{\ApJstyle j} f_{\ApJstyle i,j} E_{\ApJstyle i,j}
                \kappa_{\ApJstyle i,j}^{\ApJstyle R}
\over
\sum_{\ApJstyle j} f_{\ApJstyle i,j}E_{\ApJstyle i,j}
                                                }  \,\, , \eqno(\eqn)$$
where $\kappa_{\ApJstyle i,j}^{\ApJstyle R}$ is evaluated
at energy $E_{\ApJstyle i,j}$.
We have canceled out the frequency-independent
factors in the numerator and denominator of equation~(\showeqn -0).
Note that by the assumption of an infinite, homogeneous,
isotropic, and homogeneous medium, the $f_{\ApJstyle i,j}$'s describe
the $\gamma$-ray removal spectrum as well as emission spectrum.
Thus equation~(\showeqn -0) could also be described as
a removal-weighted mean opacity prescription.


      Equation~(\showeqn -0) has been used to construct the mean
opacities and mean fractional component opacities
for the 0th through 5th order for $^{\ApJstyle 56}$Co and 
$^{\ApJstyle 56}$Ni $\gamma$-ray fields for several cases.
The mean fractional component opacities, which are of direct use in
\S~4, are defined by
$$\xi_{\ApJstyle i}^{\ApJstyle R}=
 \kappa_{\ApJstyle i}^{\ApJstyle R}/\kappa_{\ApJstyle i}
                                         \,\, , \eqno(\eqn)$$
where $R$ is either ``a'' or ``s''.
The results are shown in Tables~I and~II.
Table~I includes the infinite order mean opacities which
are for zero energy $\gamma$-rays.
(We demonstrate in Appendix~B that infinite order Compton
or iso-Compton scattering will degrade $\gamma$-ray energy to zero.)
The $\gamma$-ray spectrum data for nuclear decays of
$^{\ApJstyle 56}$Co and $^{\ApJstyle 56}$Ni were taken from Huo
(1992) and Browne \&~Firestone (1986).
We have included in the $\gamma$-ray spectrum of
$^{\ApJstyle 56}$Co $\gamma$-rays to account for the
annihilation of the positron that is produced in $19\,$\% of 
$^{\ApJstyle 56}$Co decays (Huo 1992).
We assume that the positron is annihilated locally after a
negligible time delay.
To maintain our line spectrum assumption for the $\gamma$-rays
and
for consistency with our pair production opacity assumptions, we
assume the positron annihilates to create
two $m_{\ApJstyle e}c^{\ApJstyle 2}$ $\gamma$-rays.
A three-continuum-photon process for annihilation is of course
actually possible (see \S~1).
The X-ray spectra of the nuclear decays were not included
for the reasons discussed in \S~5.

    The emissivity-weighted mean $\gamma$-ray energy
$\bar E_{\ApJstyle i}$ for each order $i$ in Tables~I and~II 
has been calculated from the expression
$$  \bar E_{\ApJstyle i}=
{\sum_{\ApJstyle j} f_{\ApJstyle i,j}E_{\ApJstyle i,j}
\over
\sum_{\ApJstyle j} f_{\ApJstyle i,j}            }  \,\, . \eqno(\eqn)$$

     In Table~I, we show the mean opacities calculated only using
iso-Compton opacity
for the case of $\mu_{\ApJstyle e}=1$ (a fiducial case) and the case of
$\mu_{\ApJstyle e}=\mu_{\ApJstyle e}^{\ApJstyle\odot}=1.179$ which is
the mean solar system value based on the results of Anders \&~Grevesse
(1989).
There are a few things to note from Table~I.
First, that the mean $\gamma$-ray
energies and mean fractional component opacities
are the same for the two cases and that the mean opacities
in the second case are all exactly a factor of 1.179 lower than
the opacities in the first case.
This behavior occurs because $^{\ApJstyle 56}$Co
and $^{\ApJstyle 56}$Ni $\gamma$-ray spectra
(i.e., the 0th order spectra) are the
same in the two cases and the iso-Compton opacity 
differs only by an overall scale factor of 1.179 between the
two cases.
Thus for each order of scattering the two cases have the same
$\gamma$-ray spectrum shapes and generate mean opacities differing
only by 1.179.
The situation is actually clear from
equations~(\showeqn -4)--(\showeqn -0) where $\mu_{\ApJstyle e}$
cancels out when there is only iso-Compton opacity, except in
evaluation
of the mean opacities.
The optical depth of any part of the medium is larger in the 
$\mu_{\ApJstyle e}=1$ case, of course, because the absolute sizes
of the mean opacities are larger.

    Second, we note that the mean opacity and fractional
mean scattering opacity increase with order while
the fractional mean absorption opacity decreases with order.
These results occur because the iso-Compton opacity
and its fractional component opacities have the same
general behavior with decreasing energy (and
therefore increasing order) as the Compton opacity and its
fractional component opacities (see \S~2).


     Third, we note that the $\gamma$-ray mean energy
decreases most in the 1st scattering and that by the 5th order
it is only a small fraction of its 0th order value.
The decline in mean energy with increasing order is in fact
rather slow after the 1st scattering because, as noted
above,
the iso-Compton fractional absorption opacity decreases with
decreasing energy.
In the limit of infinite scattering order, the $\gamma$-rays
have lost all their energy and their mean opacity has become
pure scattering opacity:  angle-average Thomson scattering opacity
in fact (see \S~2).
Iso-Compton scattering conserves photon number and so, in principle,
no photons are destroyed or created in scattering to infinite order.

     In Table~II we present the mean opacities and
fractional mean component opacities now including
pair production and photoelectric opacities
for the 0th through 5th orders for the solar composition
(to be precise the solar system composition of Anders \&~Grevesse 1989)
and for the mean composition of SN~Ia model~W7
(Thielemann~\etal 1986).
The $\kappa_{\ApJstyle\rm pe}$ values for the
photoelectric opacity (see \S~2) were constructed from the
results given by Veigele (1973);  we did not use the scaling
approximation equation~(6).
Veigele's photoelectric cross sections extend down only to
$10^{\ApJstyle -4}\,$MeV.
This is not a problem since none of the $\gamma$-rays used
creating Table~II went below $10^{\ApJstyle -4}\,$MeV even
by the 10th scattering order.

     By comparison of the mean opacities for the $\mu_{\ApJstyle e}=1.179$
case in Table~I to those for the solar composition case in Table~II we see
that both pair production opacity
and the photoelectric opacity are nearly negligible.
This shows that for solar composition only Compton opacity 
is important through to the 5th order field.
Since higher order fields can be treated very crudely---there
is not much energy deposition to be obtained from them---it
is clear that Compton opacity alone is adequate for treating
the solar composition.

     Note that the nonzero order mean energies for the
solar composition in Table~II are actually slightly higher
than the mean energies for the
$\mu_{\ApJstyle e}=1.179$ case in Table~I despite there
being more opacity in the calculations for Table~II.
Higher mean energies are made possible by the fact that
the photoelectric opacity tends to preferentially destroy the lower
energy $\gamma$-rays and the destroyed $\gamma$-rays, of course, make
no contribution to the mean $\gamma$-ray energy.
The effect of the pair production opacity on the mean energies
can, however, either lower or raise them since pair
production opacity has both absorption and scattering components
whose relative size depends on $\gamma$-ray energy.
(Note that there is a $^{\ApJstyle 56}$Ni $\gamma$-ray with enough
energy for pair production even though the  $^{\ApJstyle 56}$Ni
0th order mean energy is much less than $2m_{\ApJstyle e}c$.)
It is the combination of the small photoelectric and pair
production opacities that results in the nonzero order
mean energies being slightly higher in the Table~II solar
composition case.

     The total scattered energy in each order (which is proportional to
$\sum_{\ApJstyle j} f_{\ApJstyle i,j}E_{\ApJstyle i,j}$)
is slightly higher for the 1st through 3rd orders for
$^{\ApJstyle 56}$Co in the solar composition case of
Table~II than for the Table~I $^{\ApJstyle 56}$Co cases.
This is because of the relatively large fractional scattering component
of the pair production opacity.
(Note only in the 0th order are there $^{\ApJstyle 56}$Co $\gamma$-rays
with enough energy to interact with the pair production opacity, but,
of course, 0th order interactions affect the higher order results.)
In higher orders for $^{\ApJstyle 56}$Co and in all nonzero
orders for $^{\ApJstyle 56}$Ni in the solar composition case,
the total scattered energy is slightly lower than in the Table~I
cases.
This is, of course, because of absorption by the photoelectric opacity.
The fraction of initial $\gamma$-ray energy remaining by the 5th order is
0.06121 for $^{\ApJstyle 56}$Co and 0.1301 for $^{\ApJstyle 56}$Ni
in the Table~I cases
and
0.06100 for $^{\ApJstyle 56}$Co and 0.1291 for $^{\ApJstyle 56}$Ni
in the Table~II solar composition case.

     The (mean) model~W7 composition is an almost all metal composition
and more than $60\,$\% of it is made of iron peak elements.
Pair production and photoelectric opacities are typically much
larger for metals than for hydrogen and helium 
(see \S~2, eq.~[2] and Veigele 1973).
Thus we expect them to be more important for the model~W7 composition
than for the solar composition.
On the other hand, $\mu_{\ApJstyle e}$ is about 1.8 times
larger for the model~W7 composition than for the solar composition.
This simply means the model~W7 composition has only about
half the electrons per unit mass and thus half the Compton opacity
that the solar composition has.
Comparing the mean total opacities and mean
fractional absorption opacities of the solar
and model~W7 compositions, it
is clear that the increased metallicity of the model~W7 composition
cannot compensate for its reduced iso-Compton opacity for
the lower orders of scattering,
but can more than compensate for the higher orders.
This result is explained by the photoelectric
opacity which grows strongly with decreasing energy
and is pure absorption opacity.

     The increased
pair production               
and photoelectric opacities   
in the model~W7 composition case relative to the solar composition
case cause the mean $\gamma$-ray energies in each nonzero order of
scattering to be slightly higher in the model~W7 composition case.
The total energy, however, in each order from the 2nd order on for
$^{\ApJstyle 56}$Co and from the 1st order on for
$^{\ApJstyle 56}$Ni is lower 
in the model~W7 composition case.
The fraction of initial $\gamma$-ray energy remaining by the 5th order
in the model~W7 composition case is
0.007632 for $^{\ApJstyle 56}$Co and 0.01025 for $^{\ApJstyle 56}$Ni.

      Pair production, despite the high metallicity,
is actually only a small contributor to the mean opacities in
the model~W7 composition case.
For $^{\ApJstyle 56}$Co in the 0th order, pair production opacity
contributes only 3$\,$\%, 2.5$\,$\%, and 6$\,$\% to the mean
total, absorption, and scattering opacities, respectively.
For $^{\ApJstyle 56}$Ni in the 0th order, pair production opacity
contributes much less than a percent to the mean opacities. 
The contribution in all other orders for
both $^{\ApJstyle 56}$Co and $^{\ApJstyle 56}$Ni is zero since there
are no $\gamma$-rays over the threshold energy of
$2m_{\ApJstyle e}c^{\ApJstyle 2}$.

     We have also calculated the emissivity-weighted inverse-mean
opacities for the 0th through 5th orders for the cases reported
in Tables~I and~II.
The differences between these and the emissivity-weighted
mean opacities tend to decrease with order.
In the 0th order they range from a few percent up to
$25\,$\% in the worst case.
The differences in the nonzero orders are quite small and,
except in the
model~W7 composition $^{\ApJstyle 56}$Ni case, are less, usually
much less, than $\sim 3\,$\%. 
Even in the model~W7 composition $^{\ApJstyle 56}$Ni case,
the differences are no more than 12$\,$\% in the 1st order
and diminish to being less $\sim 2\,$\% by the 5th order.
The reason for the reduction of the differences beyond the
0th order is the nature of the iso-Compton opacity.
It turns out that $\gamma$-ray energies in the range
$\sim 0.1$--$5\,$MeV are diminished in an iso-Compton
scattering to energies in the range $\sim 0.08$--$0.24\,$MeV.
Thus the relative variation in the energies of typical supernova decay
$\gamma$-rays is greatly reduced by the first iso-Compton scattering. 
This means that the relative variation in the opacity for
these $\gamma$-rays is greatly reduced, and thus the two kinds of 
mean opacities tend to converge.
Other kinds of mean opacities should tend to converge as well.

     The grey approximation is, obviously, always good
when the range in variation of the input opacities is
sufficiently small.
That the two kinds of mean opacities we have investigated converge
so well in the nonzero orders suggests the grey
approximation will always be good for these orders. 
This conclusion, of course, holds only insofar as the
iso-Compton opacity is a good approximation to the
Compton opacity.
The fact that in the 0th order the two kinds of mean
opacities are in not in so close agreement as in the
nonzero orders does not necessarily mean the
grey approximation is worse than in nonzero orders.
It just means that the particular mean opacity prescription
we have chosen for the grey approximation for 
supernova $\gamma$-ray transfer needs to be
adequate.
From the arguments given above, we believe that our
choice of the emissivity-weighted mean opacity
prescription has a good chance of being adequate.
 
     To end this section, we note that whenever Compton scattering
opacity is the dominant form of opacity, the mean opacities tend to
simply scale with $\mu_{\ApJstyle e}^{\ApJstyle -1}$.
This is because Compton opacity depends directly on the ratio of
electron density to mass density (see \S~2, eq.~[8]). 
Even for the model~W7 composition case where Compton
opacity is least important, the mean opacities of the
lowest orders of scattering can be obtained very roughly by
dividing the $\mu_{\ApJstyle e}=1$ mean opacities in Table~I
by 2.095.
\vskip 2\baselineskip

\centerline{4.\ \  THE LS GREY RADIATIVE TRANSFER PROCEDURE}\nobreak
\vskip\baselineskip\nobreak
     What we want from a grey radiative transfer procedure is
the $\gamma$-ray energy deposition as function of position.
We will measure this energy deposition by $\epsilon_{\ApJstyle\rm d}$,
the energy deposited per unit time per unit mass.
(Energy deposition tends to decrease with decreasing density
whether measured per mass or per volume.
Measuring per mass, however, gives a smaller range in variation.)
To get the energy deposition we assume the emissivity-weighted mean
opacities we developed in \S~3 adequately control the transfer of the
frequency-integrated radiation fields of all orders of scattering.
The energy deposition from one radioactive species is then given by
$$ \epsilon_{\ApJstyle\rm d}
 =4\pi\sum_{\ApJstyle i=0}^{\ApJstyle\infty}
   { \chi_{\ApJstyle i}^{\ApJstyle\rm a} J_{\ApJstyle i} \over \rho}
 =4\pi\sum_{\ApJstyle i=0}^{\ApJstyle\infty}
    \kappa_{\ApJstyle i}^{\ApJstyle\rm a} J_{\ApJstyle i}
                                         \,\, ,       \eqno(\eqn)$$
where $J_{\ApJstyle i}$ is the $i$th order (frequency-integrated)
mean intensity, 
$\chi_{\ApJstyle i}^{\ApJstyle\rm a}
=\kappa_{\ApJstyle i}^{\ApJstyle\rm a}\rho$ is the $i$th order
mean absorption extinction, and
$\kappa_{\ApJstyle i}^{\ApJstyle\rm a}$ is the $i$th order
mean absorption opacity.

      The 0th order mean intensity (or radiation field) is
generated by the true energy source, the radioactive species.
Formally this field at any point is given by
$$ J_{\ApJstyle 0}=\oint{d\Omega\over 4\pi}
   \int_{\ApJstyle 0}^{\ApJstyle\tau_{\ApJstyle 0}}
    dx_{\ApJstyle 0}\,S_{\ApJstyle 0}\left(x_{\ApJstyle 0}\right)
       \exp\left(-x_{\ApJstyle 0}\right)
                                \,\, ,                  \eqno(\eqn)$$ 
where $\Omega$ is solid angle,
$x_{\ApJstyle 0}$ is the optical depth measured
from the point backward along a beam
path, $\tau_{\ApJstyle 0}$ is the optical depth along the beam path
to the surface of the medium, and $S_{\ApJstyle 0}$ is the $\gamma$-ray
source function.
The source function is the radioactive species (frequency-integrated)
emissivity divided by 0th order mean total extinction:
$S_{\ApJstyle 0}=\eta_{\ApJstyle 0}^{\ApJstyle\gamma}/\chi_{\ApJstyle 0}$.
The optical depths for the 0th and all other orders are computed using
$$ dx_{\ApJstyle i}=ds\,\chi_{\ApJstyle i}
              =ds\,\kappa_{\ApJstyle i}\rho \,\, , \eqno(\eqn)$$
where $\chi_{\ApJstyle i}$ is, of course, the $i$th order mean total
extinction, $\kappa_{\ApJstyle i}$ is the $i$th order
mean total opacity, and $s$ is again the beam path length.

     Now the emissivity for any scattered radiation field $i$ is 
$\chi_{\ApJstyle i-1}^{\ApJstyle\rm s}J_{\ApJstyle i-1}$.  
Thus the source function for any scattered field $i$ is 
$\left(\kappa_{\ApJstyle i-1}^{\ApJstyle\rm s}
       /\kappa_{\ApJstyle i}\right)J_{\ApJstyle i-1}$,
where the ratio
$\kappa_{\ApJstyle i-1}^{\ApJstyle\rm s}/\kappa_{\ApJstyle i}$ is
a constant in our case.
Adapting equation~(\showeqn -1), {\it mutatis mutandis}, 
the mean intensity for $i\geq 1$ is given by
$$ J_{\ApJstyle i}=
{\kappa_{\ApJstyle i-1}^{\ApJstyle\rm s}
 \over\kappa_{\ApJstyle i}}
\oint{d\Omega\over 4\pi}
   \int_{\ApJstyle 0}^{\ApJstyle\tau_{\ApJstyle i}}
    dx_{\ApJstyle i}\,
 J_{\ApJstyle i-1}\left(x_{\ApJstyle i}\right)
       \exp\left(-x_{\ApJstyle i}\right)
                                \,\, .                  \eqno(\eqn)$$

       The 0th order radiation field must be calculated numerically.
It is in fact a straightforward integration since the 0th order
source function is known from the properties of the radioactive
species, our mean opacities, and the composition of the
supernova model being used.
We now make the sweeping assumption that all scattered radiation
fields can be calculated in the LS~approximation (see \S~1):
i.e., we assume that $J_{\ApJstyle i-1}\left(x_{\ApJstyle i}\right)$
in equation~(\showeqn -0) (which is the source function for
the $i$th field) can be replaced by its value at
$x_{\ApJstyle i}=0$ which we denote simply by $J_{\ApJstyle i-1}$.
We then pull $J_{\ApJstyle i-1}$ out of the integral in
equation~(\showeqn -0), do the integration over optical
depth, and obtain
$$ J_{\ApJstyle i}=
 J_{\ApJstyle i-1}
{\kappa_{\ApJstyle i-1}^{\ApJstyle\rm s}
 \over\kappa_{\ApJstyle i}}\zeta_{\ApJstyle i} \,\, , \eqno(\eqn)$$
where
$$\zeta_{\ApJstyle i}\equiv\oint{d\Omega\over 4\pi}\,
       \exp\left[1-\exp\left(-\tau_{\ApJstyle i}\right)\right]
                                \,\, .                  \eqno(\eqn)$$
It now follows that all scattered fields can be obtained
from
$$ J_{\ApJstyle i}=J_{\ApJstyle 0}\prod_{\ApJstyle j=1}^{\ApJstyle i}
{\kappa_{\ApJstyle j-1}^{\ApJstyle\rm s}
 \over\kappa_{\ApJstyle j}}\zeta_{\ApJstyle j} \,\, . \eqno(\eqn)$$
In equation~(\showeqn -0) and in all similar cases, we take the product
expression to be 1 when the lower limit exceeds the upper limit.

     We note here that the LS~approximation becomes good when the
emission regions
contributing significantly to the $\gamma$-ray field at a point
vary only linearly with distance in their physical state from
the conditions at the point.
Only linear variation will occur in sufficiently optically
thick conditions because the emission regions will be close to the
point.
In this case the average physical state of these regions is well
approximated by the local state at the point.
In the optically thick limit, the emission regions become the 
local region itself and the LS~approximation becomes exact.
In non-optically thick conditions the LS~approximation becomes 
a rough approximation.

      Using equation~(\showeqn -0), the expression for the energy
deposition can be written
$$ \epsilon_{\ApJstyle\rm d}
 =4\pi\kappa_{\ApJstyle\rm eff}^{\ApJstyle\rm a} J_{\ApJstyle 0}
                                               \,\, , \eqno(\eqn)$$
where
$$ \kappa_{\ApJstyle\rm eff}^{\ApJstyle\rm a}\equiv
    \sum_{\ApJstyle i=0}^{\ApJstyle\infty}
     \kappa_{\ApJstyle i}^{\ApJstyle\rm a}
      \prod_{\ApJstyle j=1}^{\ApJstyle i}
{\kappa_{\ApJstyle j-1}^{\ApJstyle\rm s}
 \over\kappa_{\ApJstyle j}}\zeta_{\ApJstyle j}  \eqno(\eqn)$$
is the effective absorption opacity.
The effective absorption opacity can be rewritten  
$$ \kappa_{\ApJstyle\rm eff}^{\ApJstyle\rm a}=
    \kappa_{\ApJstyle 0}L \,\, ,\qquad{\rm where}\qquad
L=
\sum_{\ApJstyle i=0}^{\ApJstyle\infty}
     \xi_{\ApJstyle i}^{\ApJstyle\rm a}
      \prod_{\ApJstyle j=0}^{\ApJstyle i-1}
       \xi_{\ApJstyle j}^{\ApJstyle\rm s}
        \zeta_{\ApJstyle j+1}                      \eqno(\eqn)$$
is what we call the LS approximation series and the
$\xi_{\ApJstyle i}^{\ApJstyle R}$ quantities are the mean fractional
component opacities defined by equation~(\showeqn -11) in \S~3.2.
The series $L$ is the ratio of
the energy that is ultimately absorbed from fields of all orders
to the energy removed (but not necessarily absorbed)
from the 0th order field. 

     There are three statements to be made about the series $L$ in
equation~(\showeqn -0).
First, on physical grounds alone it must converge.
Second, in the optically thin limit for nonzero scattering
orders (i.e., when $\zeta_{\ApJstyle i}=0$
for all $i\geq1$) $L$ goes to $\xi_{\ApJstyle 0}^{\ApJstyle\rm a}$, and
thus
$$ \kappa_{\ApJstyle\rm eff}^{\ApJstyle\rm a}=
             \kappa_{\ApJstyle 0}^{\ApJstyle\rm a} \,\, . \eqno(\eqn)$$
The local-state approximation becomes exact in this case for
the trivial reason that there is no energy deposited in the nonzero
orders. 
Third, if all nonzero scattering orders are in the optically
thick limit and we have an appropriate set of
order mean opacities (see Appendix~B),
$L$ converges to 1.
This means that
$$ \kappa_{\ApJstyle\rm eff}^{\ApJstyle\rm a}=
             \kappa_{\ApJstyle 0}   \eqno(\eqn)$$
and all the energy processed by the 0th order total
opacity (i.e., $4\pi\kappa_{\ApJstyle 0} J_{\ApJstyle 0}$)
is ultimately absorbed.
In Appendix~B we prove that iso-Compton opacity and Compton opacity itself
do give an appropriate set of order mean opacities for this
to happen.
The addition of a pure absorption opacity (e.g., the photoelectric
opacity) or to finite order any other opacity
(e.g., the pair production opacity) would not alter the
convergence of $L$ to 1.
Thus, for the supernova case
$\kappa_{\ApJstyle\rm eff}^{\ApJstyle\rm a}=\kappa_{\ApJstyle 0}$
in the optically thick limit of the nonzero scattering orders.
Recall also that LS~approximation becomes exact in the optically
thick limit.

     Now we could compute the terms in the $L$ series in
equation~(\showeqn -2) to any order we wish.
But computing to a high order will not necessarily yield
high accuracy because of the approximations we have made.
Instead we propose to approximate the terms in
the $L$ series so that we can evaluate it to infinite
order.
In this way we guarantee that the $L$ series will go
to the exact
nonzero order optically thick, as well as thin, limit.
In fact, in the supernova case if the nonzero orders are in
the optically thin or thick limits, then the 0th order will 
be in those limits or nearly as well because of the nature
of Compton opacity.
If all the orders are in the optically thin limit, then
grey radiative transfer with the emissivity-weighted mean
opacities is exact for supernovae (assuming time-independent,
static radiative transfer) since
it is exact for the optically thin limit 0th order
$\gamma$-ray transfer 
(assuming time-independent,
static radiative transfer)
as we showed in \S~3.1.  
If all orders are in the optically thick limit, then
for supernova $\gamma$-ray transfer
all $\gamma$-rays are locally trapped in all orders
including the 0th and thus (following from
the discussion in the last paragraph) are locally absorbed.
Thus any treatment, however crude, that ensures complete local
absorption is exact.
Therefore practically speaking for supernova $\gamma$-ray
transfer, the approximated $L$ series we propose will yield
exact results (within numerical limitations and
not counting errors due to neglecting time-dependent, non-static
radiative transfer effects) in the (all-order) optically thin and thick
limits.

      To obtain the approximated $L$ series we split the
$L$ series into low and high order terms with the $k$th term
being the first high order term:
$$ L=
\left(\sum_{\ApJstyle i=0}^{\ApJstyle k-1}
     \xi_{\ApJstyle i}^{\ApJstyle\rm a}
      \prod_{\ApJstyle j=0}^{\ApJstyle i-1}
       \xi_{\ApJstyle j}^{\ApJstyle\rm s}
        \zeta_{\ApJstyle j+1} \right)
+
\left(\prod_{\ApJstyle j=0}^{\ApJstyle k-1}
       \xi_{\ApJstyle j}^{\ApJstyle\rm s}
        \zeta_{\ApJstyle j+1}\right) 
\sum_{\ApJstyle i=k}^{\ApJstyle\infty}
     \xi_{\ApJstyle i}^{\ApJstyle\rm a}
      \prod_{\ApJstyle m=k}^{\ApJstyle i-1}
       \xi_{\ApJstyle m}^{\ApJstyle\rm s}
        \zeta_{\ApJstyle m+1}   \,\, .         \eqno(\eqn)$$
We now approximate all the high order quantities by their
$k$th values.
Then the high order terms can be summed analytically and
we obtain
$$ L=
\left(\sum_{\ApJstyle i=0}^{\ApJstyle k-1}
     \xi_{\ApJstyle i}^{\ApJstyle\rm a}
      \prod_{\ApJstyle j=0}^{\ApJstyle i-1}
       \xi_{\ApJstyle j}^{\ApJstyle\rm s}
        \zeta_{\ApJstyle j+1} \right)
+
\left(\prod_{\ApJstyle j=0}^{\ApJstyle k-1}
       \xi_{\ApJstyle j}^{\ApJstyle\rm s}
        \zeta_{\ApJstyle j+1}\right) 
{ \xi_{\ApJstyle k}^{\ApJstyle\rm a} \over
   1-\xi_{\ApJstyle k}^{\ApJstyle\rm s}
        \zeta_{\ApJstyle k} }      \,\, .         \eqno(\eqn)$$
(Note one should never choose $k$ such that 
$\xi_{\ApJstyle k}^{\ApJstyle\rm s}=1$
and $\xi_{\ApJstyle k}^{\ApJstyle\rm a}=0$.
For supernovae this is not a concern since in practice
$\xi_{\ApJstyle i}^{\ApJstyle\rm s}=1$ never happens for any
order $i$.)
In the nonzero order optically thin limit
(or in the 0th order optically thin limit if $k=0$),
$L$ reduces to $\xi_{\ApJstyle 0}^{\ApJstyle\rm a}$ of course.
In the nonzero order optically thick limit
(or in the 0th order optically thick limit if $k=0$),
we find, as desired,
$$ L=1-\prod_{\ApJstyle j=0}^{\ApJstyle k-1}
       \xi_{\ApJstyle j}^{\ApJstyle\rm s}
      +\prod_{\ApJstyle j=0}^{\ApJstyle k-1}
       \xi_{\ApJstyle j}^{\ApJstyle\rm s}
    =1                      \,\,           ,      \eqno(\eqn)$$
where we have used the fact that
$\xi_{\ApJstyle i}^{\ApJstyle\rm a}+\xi_{\ApJstyle i}^{\ApJstyle\rm s}=1$
and equation~(B5) from Appendix~B.

      The mean opacity values needed for determining
the $\zeta_{\ApJstyle i}$'s and $\xi_{\ApJstyle i}^{\ApJstyle R}$'s
in equation~(\showeqn -1) up to 5th order for
some opacity cases can be obtained
or scaled from the results shown in \S~3.2, Tables~I and~II.
The evaluation of equation~(\showeqn -1) will probably not be too
sensitive to the choice of $k$ since the amount of energy absorbed
in a scattering order decreases with scattering order as can be
seen from Tables~I and~II.  
We have computed the net energy deposition for model~W7 from early
times to day~1000 after the explosion using $k=2$ and $k=5$.
The difference between the two cases was always less than
$\sim 0.6\,$\% and it vanishes, of course, at early optically thick and
late optically thin times.
For the other opacity cases shown in Tables~I and~II we expect
similar results.
We will take $k=5$ to be our fiducial $k$ value in order to avoid
regarding $k$ as a free parameter.
The $k=5$ choice should be adequately high for all cases.

     The integration for the $\zeta_{\ApJstyle i}$'s cannot be done
exactly analytically.
For calculations in spherical symmetry, we propose using the
simple two-stream approximation
$$ \zeta_{\ApJstyle i}\approx
{1\over2}\left\{ 
\left[1-\exp\left(-\tau_{\ApJstyle i,{\rm out}}\right)\right]
+\left[1-\exp\left(-\tau_{\ApJstyle i,{\rm in}}\right)\right]\right\}
                                             \,\, ,  \eqno(\eqn)$$
where $\tau_{\ApJstyle i,{\rm out}}$ and $\tau_{\ApJstyle i,{\rm in}}$
are the outward and inward radial $\tau_{\ApJstyle i}$ values.
The calculation of the $\zeta_{\ApJstyle i}$ with the two-stream
approximation is still non-local, but it is numerically trivial.
For atmospheres without spherical symmetry some other prescription
for $\zeta_{\ApJstyle i}$ is needed.

     Having done the numerical
integration for the 0th order radiation field (which uses the
0th order total opacity $\kappa_{\ApJstyle 0}$) and
the evaluation of effective absorption opacity using
equation~(\showeqn -6) and
the $L$ series in equation~(\showeqn -2), we can then
obtain the energy deposition from equation~(\showeqn -8).
These operations together constitute the LS grey radiative transfer
procedure for obtaining the energy deposition.

      The only real elaboration of LS procedure beyond the SSH procedure
is in the calculation of effective absorption opacity.
We have discussed this calculation in detail, but the actual
coding for it is straightforward.
The basic SSH~procedure fortran code of circa 1996 (Sutherland 1996) is about
60 lines excluding comment lines and auxiliary subroutines.
The LS~procedure fortran code that we have written (starting from
the SSH code) is about 260 lines 
excluding comment lines, auxiliary subroutines, and data statements.
The comparison is not completely fair, however, since the LS~code is
more general than the SSH~code and contains some purely diagnostic lines.
The actual difference in complexity between the SSH and~LS codes is small.
Both codes run practically instantly by human perception for ordinary
supernova models. 

      Unlike the optimized SSH~procedure, there are no free parameters in
the LS~procedure.
There are, however, many significant approximations.
In \S~6, we will discuss the adequacy of some of these approximations.
\vskip 2\baselineskip

\centerline{5.\ \  THE RADIOACTIVE SOURCES}\nobreak
\vskip\baselineskip\nobreak
    In this section we review some material needed
for the treatment of the radioactive sources in the energy deposition
in supernovae.   

    The frequency-integrated emissivity in energy form $i$
for a radioactive isotope is given by
$$ \eta_{\ApJstyle 0}^{\ApJstyle i}
  ={1\over 4\pi}{n\over t_{\ApJstyle e}}
     Q_{\ApJstyle i}    \,\, , \eqno(\eqn)$$
where $Q_{\ApJstyle i}$ is the energy emitted per decay
in form $i$,
$n$ is the number density of the isotope, and
$t_{\ApJstyle e}$ is the $e$-folding time for the decay.
Since the volume of any mass element in supernova ejecta is
changing constantly, it is more convenient to express the
energy from radioactive decay in the form of energy
generation per unit time per unit mass
$$\epsilon_{\ApJstyle i}={4\pi\eta_{\ApJstyle 0}^{\ApJstyle i}
                        \over\rho}         \,\, . \eqno(\eqn)$$

     For the case of a radioactive isotope synthesized in
a supernova explosion, $\epsilon_{\ApJstyle i}$ can be expressed by 
$$\epsilon_{\ApJstyle i}=
  C_{\ApJstyle i}X(0)\exp\left(-t/t_{\ApJstyle e}\right)
                                              \,\, , \eqno(\eqn)$$
where $X(0)$ is the isotope's initial mass fraction and
$t$ is the time since explosion.
The $C$~coefficient is defined by
$$ C_{\ApJstyle i}={  Q_{\ApJstyle i}
   \over m_{\ApJstyle\rm amu} A t_{\ApJstyle e} }
                                        \,\, , \eqno(\eqn)$$
where $A$ is the isotope's atomic mass
and $m_{\ApJstyle\rm amu}$ is again the atomic mass unit.
For a radioactive isotope which is the child of an isotope synthesized
in the explosion, $\epsilon_{\ApJstyle i}$ can be expressed by
$$\epsilon_{\ApJstyle i}=
  D_{\ApJstyle i}X_{\ApJstyle 1}(0)
\left[ \exp\left(-t/t_{\ApJstyle e,2}\right)
        -\exp\left(-t/t_{\ApJstyle e,1}\right)\right]
                                             \,\, ,  \eqno(\eqn)$$
where 1 and 2 designate parent and child nucleus quantities, respectively.
The $D$~coefficient is defined by
$$ D_{\ApJstyle i}={ Q_{\ApJstyle i,2}
     \over m_{\ApJstyle\rm amu} A_{\ApJstyle 1} }
     { 1\over \left(t_{\ApJstyle e,2}-t_{\ApJstyle e,1}\right) }
                                             \,\, .  \eqno(\eqn)$$
(Note the use of $A_{\ApJstyle 1}$ in the denominator of
equation~[\showeqnsq -0] rather than
$A_{\ApJstyle 2}$ is formally correct although the expression is for
the decay of the child nucleus.
The reason is that
$X_{\ApJstyle 1}(0)\rho/\left(m_{\ApJstyle\rm amu}A_{\ApJstyle 1}\right)$,
which comes into the derivation of equation~[\showeqnsq -0],
is the initial particle density and it is conserved in weak nuclear decays.
Since the mass change is very small in weak decays, in practice
$A_{\ApJstyle 2}$ would do as well
as $A_{\ApJstyle 1}$ of course.)

      In Table~III we present the parameters needed for treatment
of the deposition of the various forms of energy. 
The main forms are $\gamma$-rays, X-rays, and the kinetic
energy of $\beta$-particles and atomic electrons (from internal
conversions and the Auger process).
The $\gamma$-ray deposition is, of course, the main subject of this
paper.
The 0th order source function for $\gamma$-rays in terms
of $\epsilon_{\ApJstyle\gamma}$ is given by
$$ S_{\ApJstyle 0}
={\epsilon_{\ApJstyle\gamma} \over 4\pi\kappa_{\ApJstyle 0} }
                                       \,\, . \eqno(\eqn)$$
Note we include in the Table~III $Q_{\ApJstyle\gamma}$'s
the $\gamma$-ray energy
from the annihilation of any positron produced in the decay.

    The radiative transfer and energy deposition of the X-rays
can be treated in a similar manner to those of the $\gamma$-rays.
X-ray energy, however, is much lower than $\gamma$-ray energy, and so
the photoelectric opacity becomes much more important for the X-rays.
Because of the great difference between the $\gamma$-ray and
X-ray energy scales and opacities,
it is not plausible to lump these radiations together in
a single grey radiative transfer procedure.
Therefore we do not treat X-ray radiative transfer and energy
deposition in this paper.
The X-rays can, in fact, be neglected at early times when their
energy contribution is negligible.
The total energy in the X-rays per decay is typically a
few kilo-electron-volts.
The X-ray contribution, however, does increase with time
because the ejecta becomes transparent to $\gamma$-rays
sooner than to X-rays.
For model~W7 we have calculated that at 300 days the
$^{\ApJstyle 56}$Co X-rays
contribute about 2$\,$\% of the $\gamma$-ray 
deposition and at 500 days, about 7$\,$\%:
$^{\ApJstyle 56}$Ni X-rays are, of course, negligible by these
late times.   
Thus, it is possible that the X-ray contribution to the energy
deposition will be significant at late times.
If this is not the case for the $^{\ApJstyle 56}$Co decay, it
may be true for the decay of longer lived radioactive
species (see \S~1).

     The energy from $\beta$-particles, positrons to be specific, is
quite important in supernovae.
The $^{\ApJstyle 56}$Ni decay produces effectively no positrons:
the upper limit on the fraction of decays leading to a positron 
emission is $<0.0013\,$\%
(Huo 1992).
The $^{\ApJstyle 56}$Co decay produces a positron 
$19\,$\% of the time.
The $\gamma$-ray energy produced on the annihilation of the
positron should be accounted for in the $^{\ApJstyle 56}$Co
$Q_{\ApJstyle\gamma}$ as we have done in Table~III
and in the mean opacities as we have done in Tables~I and~II
in \S~3.2.
There is, however, positron kinetic energy which is
mostly lost prior to annihilation:  it is about $3\,$\% 
of the $^{\ApJstyle 56}$Co $Q_{\ApJstyle\gamma}$.
As discussed in \S~1, we assume that the positrons are completely
locally trapped and so this kinetic energy is deposited locally.
With completely local deposition the positron kinetic energy 
becomes the dominant source of energy deposition in SNe~Ia
sometime in the interval 200--300~days after the explosion
because of the increasing transparency of the ejecta to
$\gamma$-rays.
For example, for model~W7 we find that the positron kinetic
energy deposition surpasses the $\gamma$-ray energy deposition
at about day~227 after the explosion.
Even without complete local trapping the positron kinetic
energy must become dominant in SNe~Ia eventually as we know
from the analysis of SN~Ia light curves and absolute spectra
(see \S~1).
Other kinds of supernovae are much denser than SNe~Ia, and so
are much more opaque to $\gamma$-rays.
The positron kinetic energy is less important for them.

     The atomic electrons released in a radioactive decay typically
have a total kinetic energy of order a few kilo-electron-volts.
Because the atomic electrons almost certainly deposit their
kinetic energy entirely locally, the atomic electrons could make a
significant contribution to energy deposition at least at late
times and from the decay of longer lived species than
$^{\ApJstyle 56}$Co (see \S~1).
\vskip 2\baselineskip

\centerline{6.\ \  APPROXIMATIONS AND COMPARISON TO SSH}\nobreak
\vskip\baselineskip\nobreak
     In this section, we discuss the adequacy of some of the
approximations we have made and
compare the LS~procedure to the SSH~procedure.

     The most obvious approximation is the use of the grey
radiative transfer itself.
SSH have shown, however, grey radiative transfer can potentially
be done very accurately for supernovae:
global deposition
errors within $\sim 2\,$\% percent and local deposition errors
of only a few percent can be achieved with the optimized SSH~procedure.
Consequently, it is how the grey radiative transfer is done that
determines the actual accuracy.

     The most significant approximation is the replacement 
of Compton opacity (which is the dominant $\gamma$-ray opacity
in supernovae) by what we have called the iso-Compton opacity
(see \S~2).
The iso-Compton opacity approximation divides Compton opacity 
into two approximate components.
The forward component causes the nearly-forward,
nearly-coherently scattered $\gamma$-rays to be exactly forward
and coherently scattered, and thus allows them to be treated
simply as part of the 0th order $\gamma$-ray field.
The iso-Compton component (which is the iso-Compton opacity
itself) creates the approximate scattered $\gamma$-rays
fields with which we need to deal.
The flux that needs to be treated as scattered and thus its
importance are greatly reduced by using the iso-Compton opacity
approximation. 
The iso-Compton opacity can be adjusted through its $g$~parameter.
We argued in \S~2 that {\it a~priori} $g=1$ seems best and the results
described below confirm that $g=1$ gives reasonable
accuracy.

     Because the 0th order field is treated numerically in
the LS~procedure and
we have chosen a mean opacity prescription (i.e., the
emissivity-weight mean opacity) favorable for the supernova case 
and we have shown that the value of the mean opacity
is not very sensitive to the exact mean opacity
prescription (see \S~3.2), we believe that the 0th order
deposition in itself is probably treated rather well.
Now much more than half the $\gamma$-ray energy is absorbed in the
0th order (see \S~3.2, Tables~I and~II).
Thus any error contribution from the nonzero order field
treatment will be quite limited
insofar as the iso-Compton opacity approximation is valid.
To be specific, error contributions from the nonzero order
treatments of $^{\ApJstyle 56}$Co and $^{\ApJstyle 56}$Ni
will be less, and probably much less, than $20\,$\% and $40\,$\%,
respectively.
In fact, $^{\ApJstyle 56}$Ni has such a short half-life
(5.9~days) that it largely decays while supernova ejecta is optically
thick and deposition is nearly entirely local.
Thus even a crude procedure can be used to obtain nearly
exact results for $^{\ApJstyle 56}$Ni in supernovae.
One can, for example, use the $^{\ApJstyle 56}$Co mean opacity values 
for $^{\ApJstyle 56}$Ni. 

     For the nonzero order field treatment we will have error
from the LS~approximation in addition to that from the
iso-Compton and grey approximations.
The error in using the grey approximation in the
nonzero scattering orders will be small again insofar as the
iso-Compton opacity approximation is valid (see \S~3.2).
Recall that the LS~approximation accounts for the
nonzero order field deposition analytically through the 
effective absorption opacity.
The LS~approximation causes the nonzero order
$\gamma$-ray transfer to be only crudely treated,
except in the optically thin and thick limits where the
LS~approximation becomes exact for supernovae as discussed in \S~4.
One particular problem with the LS~approximation is that
minima and maxima of the 0th order $\gamma$-ray field
will tend to cause under-~and overestimates
of the higher order $\gamma$-rays fields, and thus
under-~and overestimates of the energy deposition.

     There is also error in the LS~procedure from the
neglect of time-dependent and non-static radiative transfer
effects.
We discuss the size of this error at length in Appendix~A.

     How does the LS~procedure compare to the (optimized)
SSH~procedure?
As mentioned in \S~1, the SSH~procedure uses a pure absorption
mean opacity (which we will call the SSH mean opacity
and designate by $\kappa_{\ApJstyle\rm SSH}^{\ApJstyle\rm a}$)
in a numerical radiative transfer.
The SSH~procedure uses a single value for the SSH mean opacity
for a given epoch and for all radioactive species which are
treated as a single energy source.
(SSH themselves consider only $^{\ApJstyle 56}$Co
and $^{\ApJstyle 56}$Ni.)
To compare the SSH~procedure to the LS~procedure, let us
consider a system with only a single radioactive species
for the moment.
In this case, the SSH mean opacity is in effect the 0th order mean
absorption opacity of the LS~procedure (for the that
radioactive species) plus an extra 
amount of opacity that accounts for the absorption of
the scattered fields plus perhaps time-dependent and non-static
effects.
Recall from \S~2 that simply neglecting the scattering
component of opacity will tend to underestimate net absorption.
The extra absorption opacity to account for the scattered fields
is what the LS~procedure effectively obtains by treating the
scattered fields in the LS~approximation. 
The SSH~procedure obtains the extra absorption opacity
by fitting to accurate Monte Carlo calculations.
Since those Monte Carlo calculations can also include
time-dependent and non-static effects,
those effects can be incorporated into the SSH~procedure.
The optimum SSH mean opacity must be relatively large at early
times when optical depth is high and scattered fields
are most important.
As time increases and optical depth decreases, the optimum
SSH mean opacity decreases. 
In the optically thin limit, the optimum SSH mean opacity
would reduce to the 0th order absorption opacity of the
LS~procedure if no time-dependent and non-static
effects were incorporated in the SSH
procedure or if they were negligible.

     When multiple radioactive species are included in the
system, then in the LS~procedure they must be treated
individually since emissivity weighted mean opacities
cannot be calculated {\it a~priori} for multiple species with
their different time-varying abundances.
But in the SSH~procedure the single SSH mean
opacity is obtained by a fitting procedure and does not need to
be calculated {\it a~priori}.
This is what allows the radioactive species to be treated together
as a single energy source.

    That the SSH~procedure cannot in general be optimized
without comparison to more accurate calculations is probably
not a severe problem in fact.
For the model~W7 cases SSH examined, there is a range of
SSH mean opacity values which yield reasonable accuracy
for all epochs.   
This range is $\sim 0.025$--$0.03\,{\rm cm\,g^{\ApJstyle -1}}$
for $\mu_{\ApJstyle e}=2$.
For other $\mu_{\ApJstyle e}$ cases, the range should be scaled
with $\mu_{\ApJstyle e}^{\ApJstyle -1}$ (see \S~3.2.)
For the model~W7 cases examined by SSH, the error in
global deposition in using a non-optimum value that is from
the reasonable accuracy range will be less, and often much less,
than $\sim 20\,$\% (SSH).
Nevertheless the reliance of the SSH~procedure on a free
parameter is somewhat unsatisfactory.
There may be cases where the SSH~procedure with a non-optimized
SSH mean opacity yields very bad results locally or globally.
Our procedure without free parameters has the advantage that
it adapts automatically to the optical depth conditions.

      The accuracy of the LS~procedure is probably
best tested by comparison to an accurate Monte Carlo procedure.
This comparison is beyond the scope of this paper.
We have, however, compared the results for model~W7 of the LS~procedure
to those obtained using the SSH~procedure with the optimized mean
opacities determined by SSH. 
Those optimized mean opacities did incorporate some 
time-dependent and non-static effects (Sutherland 1998).
The only radioactive sources considered are $^{\ApJstyle 56}$Co
and $^{\ApJstyle 56}$Ni.
For the comparison we have not used exactly the W7~mean opacities
from Table~II in \S~3.2.
Because of differences from SSH in $\gamma$-ray opacity data and in
the versions of model~W7 used, those values are not
quite consistent with the counterpart values used by
SSH. 
Therefore we slightly adjusted our mean opacities for model~W7 to
force consistency with SSH.
For example, instead of using the $^{\ApJstyle 56}$Co
$\kappa_{\ApJstyle 0}^{\ApJstyle\rm a}=0.0249
\,{\rm cm^{\ApJstyle 2}\,g^{\ApJstyle -1}}$
(which is derivable from Table~II),
we use
$\kappa_{\ApJstyle 0}^{\ApJstyle\rm a}=0.0255
\,{\rm cm^{\ApJstyle 2}\,g^{\ApJstyle -1}}$
for $^{\ApJstyle 56}$Co in order
to agree with the optically thin limit
optimum $\kappa_{\ApJstyle\rm SSH}^{\ApJstyle\rm a}=0.0255
\,{\rm cm^{\ApJstyle 2}\,g^{\ApJstyle -1}}$ found by SSH.
(Note the SSH $\kappa_{\ApJstyle\rm SSH}^{\ApJstyle\rm a}=0.0255 
\,{\rm cm^{\ApJstyle 2}\,g^{\ApJstyle -1}}$
value seems to have been negligibly affected by
any time-dependent, non-static effects.)
It should be emphasized that exact consistency with the
SSH parameters has not been obtained and recalled that the SSH
results themselves are not globally accurate to better than
$\sim 2\,$\% relative to their Monte Carlo results.
We judge that global discrepancies of less than $2\,$\% from the
SSH results to be negligible. 

     In Figure~1 we show the $\gamma$-ray energy deposition functions
for model~W7 calculated for day 110 after explosion using the
LS~procedure and the SSH~procedure with three different values of
$\kappa_{\ApJstyle\rm SSH}^{\ApJstyle\rm a}$.
By day~110 the $^{\ApJstyle 56}$Ni is virtually all gone and the
energy source is almost entirely $^{\ApJstyle 56}$Co.
The deposition function (which differs from the deposition
function used by SSH) is energy deposited per unit mass divided by
the mean radioactive
energy generated per unit mass for the whole model.
The integral of the deposition function with respect to mass fraction
(which we call the net deposition) is the ratio of total
energy deposited to total energy generated.

     The table on the figure identifies the calculation and gives
the calculation net deposition, $\gamma$-ray optical depth
(the $^{\ApJstyle 56}$Co effective absorption opacity optical depth
in the LS case and the SSH mean opacity optical depth in the SSH cases),
and,
for the SSH calculations, the $\kappa_{\ApJstyle\rm SSH}^{\ApJstyle\rm a}$
value.
For the LS~procedure the effective absorption opacities
(the $\kappa_{\ApJstyle\rm eff}^{\ApJstyle\rm a}$'s) are variables
of course, and so we only show the 
$^{\ApJstyle 56}$Co $\kappa_{\ApJstyle 0}^{\ApJstyle\rm a}$
which is the lower limit of the
$^{\ApJstyle 56}$Co $\kappa_{\ApJstyle\rm eff}^{\ApJstyle\rm a}$.
The Thomson optical depth shown in the figure counts all electrons,
free and bound.
Although the Thomson optical depth is not a directly
relevant physical quantity, it is a useful characteristic of the model. 

    The optimum $\kappa_{\ApJstyle\rm SSH}^{\ApJstyle\rm a}$ value 
for day~110 is $0.0264\,{\rm cm^{\ApJstyle 2}\,g^{\ApJstyle -1}}$.
The deposition function for this value is in fairly close
agreement with the LS~procedure deposition function.
The LS~procedure net deposition $0.1300$ is about 4$\,$\%
larger than the optimum SSH~procedure net deposition $0.1247$.
This $4\,$\% discrepancy is in fact the maximum discrepancy from the
optimized SSH~procedure for all epochs.
If we add this $4\,$\% discrepancy to the maximum error in
the SSH results of $\sim 2\,$\%, we obtain an estimate of
the maximum error in the
LS~procedure net deposition of $\sim 6\,$\%.
This result is derived only from model~W7 results, but we will take
it as a general estimate of the maximum error in the LS~procedure.
It may well be an underestimate:  we are not able to make the
LS~procedure parameters exactly consistent with the counterpart SSH
procedure parameters and the SSH optimization does not seem to have 
incorporated all time-dependent, non-static effects.
From the discussion in Appendix~A, a high estimate of the maximum
error in the LS~procedure is of order $20\,$\%.

     At days~50,        150,     200,     and 300 the deviations of
the LS~procedure results from the SSH~procedure results are
             $-0.2\,$\%, $3\,$\%, $2\,$\%, and $-0.3\,$\%,
respectively.
At times earlier than day~50 and later than day~300, the
discrepancies (i.e., the absolute values of the deviations)
are always less than $2\,$\%, and thus
we judge them to be negligible.
This is not unexpected given our arguments that the LS~procedure
should have negligible error in the optically thin and thick limits
(see \S~4).
The fairly good agreement with the optimized SSH~procedure setting
$g=1$ for the iso-Compton opacity (see \S~2) suggests that
$g=1$ may be generally good.
Values for $g$ significantly different from 1 did less well.
We did not search for an optimum $g$ value, however, since that value
could only be system-specific.

     To test the range sensitivity of the SSH~procedure to 
$\kappa_{\ApJstyle\rm SSH}^{\ApJstyle\rm a}$, we have also calculated
for day~110
the SSH deposition function for
$\kappa_{\ApJstyle\rm SSH}^{\ApJstyle\rm a}
 =0.0255\,{\rm cm^{\ApJstyle 2}\,g^{\ApJstyle -1}}$ which
is the optically thin limit value
and for
$\kappa_{\ApJstyle\rm SSH}^{\ApJstyle\rm a}
 =0.0277\,{\rm cm^{\ApJstyle 2}\,g^{\ApJstyle -1}}$
(more precisely $0.02769\,{\rm cm^{\ApJstyle 2}\,g^{\ApJstyle -1}}$)
which yields the same net deposition as the LS~procedure.
We can see that the $0.0255\,{\rm cm^{\ApJstyle 2}\,g^{\ApJstyle -1}}$
value yields a deposition function that is rather close to the result
for the optimum
$0.0264\,{\rm cm^{\ApJstyle 2}\,g^{\ApJstyle -1}}$ value.
The $0.0277\,{\rm cm^{\ApJstyle 2}\,g^{\ApJstyle -1}}$ value yields
a deposition function that is in close, but not
perfect, agreement with the LS~procedure deposition function.
The $0.0277\,{\rm cm^{\ApJstyle 2}\,g^{\ApJstyle -1}}$ value is
optimum for about day~75.

    Both the $0.0255\,{\rm cm^{\ApJstyle 2}\,g^{\ApJstyle -1}}$ and
the $0.0277\,{\rm cm^{\ApJstyle 2}\,g^{\ApJstyle -1}}$ values are
within the reasonable accuracy range of values for the SSH~procedure
for model~W7.
We have calculated the net deposition with the SSH~procedure
for the 
$0.0255\,{\rm cm^{\ApJstyle 2}\,g^{\ApJstyle -1}}$ and
$0.0277\,{\rm cm^{\ApJstyle 2}\,g^{\ApJstyle -1}}$ values
for the first 1000~days after explosion and compared the
results to the results obtained with the optimized
SSH mean opacities.
The $0.0255\,{\rm cm^{\ApJstyle 2}\,g^{\ApJstyle -1}}$ value
gives a maximum discrepancy of $\sim 9\,$\% (from a
deviation of $\sim -9\,$\%) at about day~35;
at late times (i.e., after day~500) when the ejecta is optically
thin the discrepancy for
the $0.0255\,{\rm cm^{\ApJstyle 2}\,g^{\ApJstyle -1}}$ value
vanishes.
The $0.0255\,{\rm cm^{\ApJstyle 2}\,g^{\ApJstyle -1}}$ value,
of course, underestimates the net deposition, except at late
times.
The $0.0277\,{\rm cm^{\ApJstyle 2}\,g^{\ApJstyle -1}}$ value's
maximum deviation is $\sim 9\,$\% at about day~1000;  this
is the asymptotic limit deviation in fact as time goes to
infinity.
Its deviation vanishes at about day~75 and is negative
at earlier times with a minimum of $\sim -6\,$\% at about day~30.

     Given the foregoing discussion, it is likely that the
LS~procedure offers at least a modest improvement in accuracy
over the unoptimized SSH~procedure.
Since finding the optimum SSH mean opacity for any particular
system requires doing the computationally intensive $\gamma$-transfer
(usually by means of a Monte Carlo calculation) that one wishes
to avoid by doing a simplified $\gamma$-ray deposition calculation,
the LS~procedure may be the best choice for that simplified
$\gamma$-ray deposition calculation. 
\vskip 2\baselineskip

\centerline{7.\ \  CONCLUSIONS}\nobreak
\vskip\baselineskip\nobreak
      We have developed a simplified, grey radiative
transfer procedure sans free parameters
for energy deposition in supernovae.
This procedure does numerical radiative transfer to handle the
0th order $\gamma$-ray field (i.e., the unscattered
$\gamma$-ray field direct from the nuclear decay) and
uses the local-state (LS)~approximation for treating the higher
order (i.e., scattered and multiply scattered) $\gamma$-ray
fields.
Because we rely on the LS~approximation we call the
procedure the LS grey radiative transfer procedure
or LS~procedure for short.
In determining the scattered fields we also rely on an approximation
of the Compton opacity which we call the iso-Compton opacity.
The parameters needed for an LS~procedure calculation for
radioactive $^{\ApJstyle 56}$Co and $^{\ApJstyle 56}$Ni
for a range of composition cases are given in 
Tables~I and~II (\S~3.2), and Table~III (\S~5).

     Probably the best test for the LS~procedure would be a comparison to
an accurate Monte Carlo procedure.
Such a test, however, is beyond the scope of this paper.
We have done a comparison to the simplified, grey radiative transfer
procedure of Swartz, Sutherland, \&~Harkness
(1995) (i.e., the SSH~procedure) which requires the adjustment
of a free parameter to obtain optimum results. 
This comparison suggests that the LS~procedure will be
modestly more reliable overall than the unoptimized
SSH~procedure.
The comparison also suggests that the maximum
error in an LS~procedure result for net deposition could
be as low $\sim 6\,$\%.
An examination of the time-dependent, non-static effects
on radiative transfer, however, suggests the maximum error
could be as high as of order $20\,$\% (see Appendix~A).
For the present, we estimate the maximum error in 
an LS procedure calculation to be of order $10\,$\%.
Only an extensive comparison to a truly high accuracy
$\gamma$-ray transfer procedure can definitively determine
the actual maximum error in using the LS~procedure.
Such a comparison is left for future work.

     Since finding the optimum SSH mean opacity requires doing
the detailed (e.g., Monte Carlo) radiative transfer one wants to avoid in
using a simplified $\gamma$-ray energy deposition procedure,
the LS~procedure may be the best choice for that simplified 
procedure.
The extra effort in developing and running an LS~procedure code
beyond that of an SSH procedure code is small.

     The LS~procedure code used for this paper can be obtained by request
from the author. 
\vskip 2\baselineskip

   This work was supported
   by the U.S.~Department of Energy's
   Office of Fusion Energy Sciences
   under Contract No.~DE-AC05-96OR22464 with
   Lockheed Martin Energy Research Corp.,
   by the ORNL Research Associates Program administered jointly
   by ORNL and the Oak Ridge Institute for Science and Education,
   and by the University of Nevada, Las Vegas.
   I thank Peter Sutherland for providing me with his SSH
   grey radiative transfer code for $\gamma$-ray energy deposition
   and David Schultz and other colleagues for their suggestions.
\vskip 2\baselineskip

\counteqn=0
\centerline{APPENDIX A}\nobreak
\vskip\baselineskip\nobreak
\centerline{TIME-DEPENDENT, NON-STATIC RADIATIVE TRANSFER}\nobreak
\vskip\baselineskip\nobreak
     In this appendix, we discuss the errors arising from the
neglect in the LS~procedure of the effects of time-dependent,
non-static radiative transfer.
Before doing so we need to specify the supernova velocity field.
This field a day at most after explosion is
ordinarily homologous expansion where the matter elements move with
a range of constant velocities and were effectively at a point at
time $t=0$, the time of explosion.
The radius of a matter element at any time $t$ after $t=0$ is given by
$$  r=vt                       \,\,  ,      \eqno({\rm A}\eqn)$$
where $v$ is the matter element's velocity.
Thus velocity can be used as a comoving coordinate for homologous
expansion.

    The characteristic velocity ``radius'', $v_{\ApJstyle\rm ch}$,
of significant $\gamma$-ray energy deposition in SNe~Ia is
$\sim 10^{\ApJstyle 9}\,{\rm cm\,s^{\ApJstyle -1}}$.
Other types of supernovae probably have smaller deposition radii
by a factor of 2 or 3. 
We will assume 
$v_{\ApJstyle\rm ch}=10^{\ApJstyle 9}\,{\rm cm\,s^{\ApJstyle -1}}$
below.

     Let us first consider time dependence.
The time-dependent effects we will consider do not seem to have
been included in SSH's Monte Carlo calculations, and thus the errors
we discuss are in addition errors relative to those in the optimized
SSH~procedure.

     Since most of the $\gamma$-ray energy is lost in the
first non-forward
Compton scattering (i.e., the first iso-Compton ``scattering''), the
characteristic distance for energy loss is the smaller of the
mean free path for the first iso-Compton ``scattering'' or the characteristic
deposition spatial radius (i.e., $v_{\ApJstyle\rm ch}t$).
Thus the characteristic lifetime
of a $\gamma$-ray, $\Delta t$, will satisfy inequality
$$ {\Delta t}\lapprox {v_{\ApJstyle\rm ch}t\over c}
       \approx0.03\times t
                                         \,\, . \eqno({\rm A}\eqn)$$
Thus,
$$ {{\Delta t}\over t}\lapprox{v_{\ApJstyle\rm ch}\over c}\approx0.03
                                         \,\, . \eqno({\rm A}\eqn)$$

     In homologous expansion the density at any velocity is proportional
to $t^{\ApJstyle -3}$, and thus optical depth is approximately
proportional to $t^{\ApJstyle -2}$.
Therefore the change in the characteristic $\gamma$-ray optical
depth $\Delta\tau$ during a $\gamma$-ray lifetime is given by 
$$  { \Delta\tau \over \tau }\approx 2{{\Delta t}\over t}
                                       \,\, . \eqno({\rm A}\eqn)$$
Now global deposition goes roughly as $1-\exp(-\tau)$.
The relative change in global deposition due to a change in
$\tau$ of $\Delta\tau$ due in turn to a change in $t$ of
$\Delta t$ is of order
$$          {\exp(-\tau) \Delta\tau \over 1-\exp(-\tau) }
         ={\Delta\tau \over \exp(\tau)-1 }
         \leq {\Delta\tau \over\tau }
                                       \,\, , \eqno({\rm A}\eqn)$$
where the equality holds only asymptotically as $\tau$ is goes to 0.
Consequently, the characteristic relative error in energy deposition
$\Delta\epsilon_{\ApJstyle\rm d}/\epsilon_{\ApJstyle\rm d}$
due to neglecting expansion during a
characteristic $\gamma$-ray lifetime will satisfy 
$$ {\Delta\epsilon_{\ApJstyle\rm d}\over\epsilon_{\ApJstyle\rm d}}
\lapprox
{ \Delta\tau \over \tau }\approx 2{{\Delta t}\over t}
\lapprox{2v_{\ApJstyle\rm ch}\over c}\approx0.06
                                         \,\, . \eqno({\rm A}\eqn)$$
We see that this error is limited and
is less than or approximately equal to the estimated maximum
error $\sim 6\,$\% (see \S~6) in the net deposition in the LS~procedure
relative to the optimized SSH~procedure.

     Besides the time since explosion $t$, there is another important
time scale relevant to time dependence:  the $e$-folding time
$t_{\ApJstyle e}$ of the radioactive decay.
If $\Delta t$ becomes comparable to $t_{\ApJstyle e}$, then deposition
at some time corresponds the radioactive decay energy generation at
a significantly earlier time.
Since the rate of energy generation goes as
$\exp\left(-t/t_{\ApJstyle e}\right)$ at least approximately
(see \S~5, eqs.~[65] and~[67]),
it follows, using equation~(A\showeqn -4), that the characteristic
relative error in deposition from neglecting the
changing rate of energy generation will satisfy
$$ {\Delta\epsilon_{\ApJstyle\rm d}\over\epsilon_{\ApJstyle\rm d}}
    \approx{\Delta t\over t_{\ApJstyle e}}\lapprox
            {v_{\ApJstyle\rm ch}\over c}
            {t\over t_{\ApJstyle e}}
                                         \,\, . \eqno({\rm A}\eqn)$$
For the case of $^{\ApJstyle 56}$Ni with 
$t_{\ApJstyle e}=8.5$~days (see \S~5, Table~III), we find
$$ {\Delta\epsilon_{\ApJstyle\rm d}\over\epsilon_{\ApJstyle\rm d}}
\lapprox 0.004\times t_{\ApJstyle\rm d}   
                                         \,\, , \eqno({\rm A}\eqn)$$
where $t_{\ApJstyle\rm d}$ is the time since explosion $t$ measured
in units of days.
By day~20 when almost all the $^{\ApJstyle 56}$Ni is gone,
${\Delta\epsilon_{\ApJstyle\rm d}/\epsilon_{\ApJstyle\rm d}}
\lapprox 0.1$, and thus time independence
should be a good approximation for the $^{\ApJstyle 56}$Ni $\gamma$-ray
deposition.
In fact since the ejecta are still optically thick to $\gamma$-rays
on day~20, the $^{\ApJstyle 56}$Ni
${\Delta\epsilon_{\ApJstyle\rm d}/\epsilon_{\ApJstyle\rm d}}$
will be much less than $0.1$.
For model~W7, the $^{\ApJstyle 56}$Ni
${\Delta\epsilon_{\ApJstyle\rm d}/\epsilon_{\ApJstyle\rm d}}$
is of order $10^{\ApJstyle -2}$ on day~20.
For supernovae other than SNe~Ia,
${\Delta\epsilon_{\ApJstyle\rm d}/\epsilon_{\ApJstyle\rm d}}$
will be smaller still on day~20 since these supernova are much
denser, have larger optical depths, and smaller $\gamma$-ray
lifetimes.
We see that for the period when $^{\ApJstyle 56}$Ni is a significant
energy source, the error in energy deposition due to neglect of the 
time variation of the $^{\ApJstyle 56}$Ni energy generation rate
will be quite small.
 
     For the case of $^{\ApJstyle 56}$Co with 
$t_{\ApJstyle e}=111.48$~days (see \S~5, Table~III), we find
$$ {\Delta\epsilon_{\ApJstyle\rm d}\over\epsilon_{\ApJstyle\rm d}}
    \lapprox 3\times10^{\ApJstyle -4} \times t_{\ApJstyle\rm d}
                                         \,\, . \eqno({\rm A}\eqn)$$
The equality version of equation~(A\showeqn -0) holds for SNe~Ia
by about 100~days or less after explosion and for other kinds
of supernovae
(which are denser and have larger optical depths at comparable
epochs) by perhaps 1000~days after explosion. 
Thus by of order 1000~days the ratio
${\Delta\epsilon_{\ApJstyle\rm d}/\epsilon_{\ApJstyle\rm d}}$
is starting to become large for any kind of supernova.
We cannot expect time-independent approximation for deposition
from $^{\ApJstyle 56}$Co
decay to hold as late as day~1000.
In the case of SNe~Ia, at 300 days after the explosion
the error due to neglecting the time variation in energy generation
could be as large as $\sim 10\,$\%. 

     Now we turn to non-static effects.
These effects were at least partially accounted for in the
optimized SSH~procedure (Sutherland 1998).
Consider the specific intensity absorbed at some point from
a beam originating some distance away.
Let $1$ designate the frame of absorption and $0$ the frame of
emission.
The beam in the frame of emission has specific intensity
$I_{\ApJstyle 0}\left(\nu_{\ApJstyle 0}\right)$ and band width
$d\nu_{\ApJstyle 0}$.
First assume a static case.
The energy absorbed (per unit volume per unit time per unit
solid angle) from the beam is
$$ \chi_{\ApJstyle 1}\left(\nu_{\ApJstyle 1}\right)
    I_{\ApJstyle 1}\left(\nu_{\ApJstyle 1}\right)\,d\nu_{\ApJstyle 1}
   =
  \chi_{\ApJstyle 1}\left(\nu_{\ApJstyle 1}\right)
    I_{\ApJstyle 0}\left(\nu_{\ApJstyle 0}\right)\,d\nu_{\ApJstyle 0}
                                         \,\, , \eqno({\rm A}\eqn)$$
where the specific intensity, frequency, and band width in frame~1 are the
equal to those in frame~0 because of the static condition.
If we now assume that the originating point of the beam is moving
away from the absorption point with velocity $\beta$ (measured in
units of $c$), then using the relativistic transformations
(e.g., Mihalas 1978, p.~495) the energy absorbed is
$$ \chi_{\ApJstyle 1}\left(\nu_{\ApJstyle 1}'\right)
    I_{\ApJstyle 1}\left(\nu_{\ApJstyle 1}'\right)\,d\nu_{\ApJstyle 1}'
   =
  \chi_{\ApJstyle 1}\left(\nu_{\ApJstyle 1}'\right)
    I_{\ApJstyle 0}
    \left(\nu_{\ApJstyle 0}\right)\,d\nu_{\ApJstyle 0}\psi^{\ApJstyle 4}
                                         \,\, , \eqno({\rm A}\eqn)$$
where
$$     \nu_{\ApJstyle 1}'=\nu_{\ApJstyle 0}\psi  \,\, , \qquad 
       d\nu_{\ApJstyle 1}'=d\nu_{\ApJstyle 0}\psi  \,\, ,
      \qquad{\rm and}\qquad
      \psi=\gamma\left(1-\beta\right)    \,\, .    \eqno({\rm A}\eqn)$$
The relativistic transformations account for the energy changes due to
the Doppler shift, 
advection,         
and aberration.    
If we now assume that extinction $\chi_{\ApJstyle 1}$ is
frequency-independent, then
the energy absorbed in the moving case is reduced by the
relative amount $\left|\psi^{\ApJstyle 4}-1\right|$ from the energy
absorbed in the static case.


     For supernovae in the optically thin limit, the appropriate
characteristic $\beta$ value is just obtained from the characteristic
velocity radius of deposition:  thus
$\beta=v_{\ApJstyle\rm ch}/c\approx0.033$.
Therefore the characteristic relative relative error in
absorbed energy in the optically thin limit due to neglecting expansion 
is
$$    \left|\psi^{\ApJstyle 4}-1\right|\approx 4\beta\approx0.13 
                                      \,\, .  \eqno({\rm A}\eqn)$$
In optically thick cases the energy loss is smaller because
the $\gamma$-rays traverse smaller velocity shifts before absorption.

    Now in general the extinction will not be frequency-independent 
(i.e.,
$\chi_{\ApJstyle 1}\left(\nu_{\ApJstyle 1}'\right)
\neq\chi_{\ApJstyle 1}\left(\nu_{\ApJstyle 1}\right)$ in general),
and thus there is another possible error in energy absorbed 
due to neglecting expansion.
However, the effective absorption opacity for the first scattering
of the decay $\gamma$-rays in supernovae is actually fairly constant.
We estimate that the Doppler shift for $\beta=0.033$
increases effective absorption opacity and hence energy
absorption by only of order $1\,$\%.
Therefore the error in energy deposition due to the frame
transformation (see eq.~[A\showeqnsq -0]) is much more
important than the error due to the Doppler shift's effect on
extinction. 

     From the foregoing analysis we can see there that could be errors
in energy deposition
of up to of order tens of percent from neglecting
the effects of time-dependent, non-static radiative transfer.
The errors tend to be largest in the optically thin epoch.
Some of these errors may cancel.
If we add in a root-mean-square sense characteristic high values
of the errors we have discussed
for cases where the LS~procedure can reasonably be used,
then the result is an overall error of order $20\,$\%.
In \S~6 we concluded for model~W7 that the maximum error in
the LS~procedure net deposition relative to the Monte Carlo
results of SSH was $\sim 6\,$\%.
This, however, was a limited test and it is not clear that
all the effects of time-dependent, non-static radiative transfer
were included in the SSH calculations.
Thus, errors larger than $\sim 6\,$\% are possible in the LS~procedure.
For now we will adopt $6\,$\% and $20\,$\% as the low and high
estimates, respectively, of the maximum error of the LS~procedure.
More extensive testing of the LS~procedure is needed to
determine its actual accuracy.
\vskip 2\baselineskip

\counteqn=0
\centerline{APPENDIX B}\nobreak
\vskip\baselineskip\nobreak
\centerline{MATHEMATICAL BEHAVIOR}\nobreak
\centerline{OF THE LS APPROXIMATION SERIES}\nobreak
\vskip\baselineskip\nobreak
     In \S~4, we derived the following series (used in evaluating the
effective absorption opacity) from the LS~approximation:
$$ L=
    \sum_{\ApJstyle i=0}^{\ApJstyle\infty}
     \xi_{\ApJstyle i}^{\ApJstyle\rm a}
      \prod_{\ApJstyle j=0}^{\ApJstyle i-1}
       \xi_{\ApJstyle j}^{\ApJstyle\rm s}
        \zeta_{\ApJstyle j+1}                      \eqno({\rm B}\eqn)$$
(see eq.~[56]).
The series $L$ is the ratio of
the energy that is ultimately absorbed from fields of all orders
to the energy removed (but not necessarily absorbed)
from the 0th order field.
In the optically thick limit for all orders $i\geq1$,
$\zeta_{\ApJstyle i\geq1}=1$ and the series reduces to
$$ L_{\ApJstyle\rm thick}=
    \sum_{\ApJstyle i=0}^{\ApJstyle\infty}
     \xi_{\ApJstyle i}^{\ApJstyle\rm a}
      \prod_{\ApJstyle j=0}^{\ApJstyle i-1}
       \xi_{\ApJstyle j}^{\ApJstyle\rm s}
                                            \,\, . \eqno({\rm B}\eqn)$$
Here we will derive some of the properties of equation~(B\showeqn -0).

     First, we define the finite series
$$ L_{\ApJstyle{\rm thick}, n}=
    \sum_{\ApJstyle i=0}^{\ApJstyle n}
     \xi_{\ApJstyle i}^{\ApJstyle\rm a}
      \prod_{\ApJstyle j=0}^{\ApJstyle i-1}
       \xi_{\ApJstyle j}^{\ApJstyle\rm s}
                                            \,\, . \eqno({\rm B}\eqn)$$
Using the relation
$$ \xi_{\ApJstyle i}^{\ApJstyle\rm a}=1-\xi_{\ApJstyle i}^{\ApJstyle\rm s}
                                                    \eqno({\rm B}\eqn)$$
it is straightforward to show that equation~(B\showeqn -1) can be rewritten
$$ L_{\ApJstyle{\rm thick}, n}=1-\prod_{\ApJstyle j=0}^{\ApJstyle n}
       \xi_{\ApJstyle j}^{\ApJstyle\rm s}
                                            \,\, . \eqno({\rm B}\eqn)$$
Now, of course,
$$ L_{\ApJstyle\rm thick}=1-\prod_{\ApJstyle j=0}^{\ApJstyle\infty}
       \xi_{\ApJstyle j}^{\ApJstyle\rm s}
                                            \,\, . \eqno({\rm B}\eqn)$$
If all $\xi_{\ApJstyle j}^{\ApJstyle\rm s}=1$, then
$$L_{\ApJstyle\rm thick}=0
                                            \,\, . \eqno({\rm B}\eqn)$$
Physically, this means that there is no absorption since the opacity
in all orders is pure scattering opacity.
Thus no flux is absorbed at all.
If there is an order $\ell$ for which $\xi_{\ApJstyle\ell}^{\ApJstyle\rm s}=0$
(implying $\xi_{\ApJstyle\ell}^{\ApJstyle\rm a}=1$), then the 
equation~(B\showeqn -5) series truncates at term $\ell$ and
equation~(B\showeqn -1) shows that
$$              L_{\ApJstyle\rm thick}=1         \,\, . \eqno({\rm B}\eqn)$$
Physically, this means no $(\ell+1)$th or higher order field can
be created because no flux is scattered by the $\ell$th order opacity
and that because of the optical thickness all the flux removed from
the 0th order field by the 0th order opacity is absorbed in a
finite number of orders (i.e., in the 0th through $\ell$th orders).

     Now consider the case that $\xi_{\ApJstyle j}^{\ApJstyle\rm s}$
satisfies $0<\xi_{\ApJstyle j}^{\ApJstyle\rm s}\leq 1$.
If
$$ \lim_{\ApJstyle j\to\infty}\xi_{\ApJstyle j}^{\ApJstyle\rm s}<1
                                            \,\, , \eqno({\rm B}\eqn)$$
then
$$\prod_{\ApJstyle j=0}^{\ApJstyle\infty}\xi_{\ApJstyle j}^{\ApJstyle\rm s}
   =0                                             \eqno({\rm B}\eqn)$$ 
(e.g., Arfken 1970, p.~286),
$$              L_{\ApJstyle\rm thick}=1 \,\, ,    \eqno({\rm B}\eqn)$$
and all the flux is absorbed in the limit
of infinite scattering.
On the other hand, if
$$ \lim_{\ApJstyle j\to\infty}\xi_{\ApJstyle j}^{\ApJstyle\rm s}=1
                                            \,\, , \eqno({\rm B}\eqn)$$
then
$$\prod_{\ApJstyle j=0}^{\ApJstyle\infty}\xi_{\ApJstyle j}^{\ApJstyle\rm s}
     \geq 0                             \eqno({\rm B}\eqn)$$ 
(e.g., Arfken 1970, p.~286),
$$        L_{\ApJstyle\rm thick}\leq 1   \,\, ,   \eqno({\rm B}\eqn)$$
and all the flux may or may not be absorbed.

     The optically thick limit LS approximation series for pure Compton
scattering is of particular interest to us.
Is $L_{\ApJstyle\rm thick}$ equal to 1 or not?
Let us assume that a Compton scattering results in a unique energy
photon (rather than in a continuum of energies dependent on the
scattering angle) and that
$\xi_{\ApJstyle j}^{\ApJstyle\rm s}=1$ only for zero photon energy. 
These assumptions are satisfied both by the
our iso-Compton opacity and the 
angle-averaged Compton opacity (see \S~2).
We will examine the degradation of a single photon in the limit
of infinite scattering.

     First consider the case that in the limit of infinite scattering
the photon energy $\alpha$ (in units of $m_{\ApJstyle e}c^{\ApJstyle 2}$)
is degraded to $\alpha_{\ApJstyle\infty}>0$.
We then have $\lim_{\ApJstyle j\to\infty}
\xi_{\ApJstyle j}^{\ApJstyle\rm s}<1$
since $\xi_{\ApJstyle j}^{\ApJstyle\rm s}=1$ only for $\alpha=0$
for both the iso-Compton opacity and
the angle-averaged Compton opacity (\S~2). 
Thus
$\prod_{\ApJstyle j=0}^{\ApJstyle\infty}\xi_{\ApJstyle j}^{\ApJstyle\rm s}
   =0$
and $L_{\ApJstyle\rm thick}=1$.
Now the product
$\prod_{\ApJstyle j=0}^{\ApJstyle\infty}\xi_{\ApJstyle j}^{\ApJstyle\rm s}$
is actually the amount of energy remaining with the photon in the
limit of infinite scattering.
If
$\prod_{\ApJstyle j=0}^{\ApJstyle\infty}\xi_{\ApJstyle j}^{\ApJstyle\rm s}
=0$, then $\alpha_{\ApJstyle\infty}=0$.
Therefore there is an inconsistency showing that our premise that
$\alpha_{\ApJstyle\infty}>0$ is incorrect.

    Taking $\alpha_{\ApJstyle\infty}=0$ implies that
$\lim_{\ApJstyle j\to\infty}\xi_{\ApJstyle j}^{\ApJstyle\rm s}=1$
which in turn implies 
$\prod_{\ApJstyle j=0}^{\ApJstyle\infty}\xi_{\ApJstyle j}^{\ApJstyle\rm s}
   \geq 0$.
However, $\alpha_{\ApJstyle\infty}=0$ directly implies that
$\prod_{\ApJstyle j=0}^{\ApJstyle\infty}\xi_{\ApJstyle j}^{\ApJstyle\rm s}
  =0$.
Thus there is consistency and we conclude that optically
thick limit iso-Compton opacity and 
angle-averaged Compton opacity do yield
$\alpha_{\ApJstyle\infty}=0$ and $L_{\ApJstyle\rm thick}=1$.
Since these approximations to the true angle-dependent
Compton opacity are consistent with the ``average''
Compton opacity behavior, we further conclude that in the
optically thick limit the true angle-dependent Compton opacity yields
$\alpha_{\ApJstyle\infty}=0$ and $L_{\ApJstyle\rm thick}=1$.
\vfill\eject

{  
\specialpage
\hoffset=.5truein                
\hsize=6.truein                  
\vskip\baselineskip
\vglue -.8truein

\centerline{TABLES}

\vskip\baselineskip

{
\baselineskip=20pt

\baselineskip=20pt
\tabskip=50pt minus 50pt  
\newdimen\digitwidth
\setbox0=\hbox{\rm0}
\digitwidth=\wd0
\catcode\lq?=\active
\def?{\kern\digitwidth}

\centerline{TABLE I}
\centerline{Mean opacities and mean fractional component opacities}
\centerline{including only iso-Compton opacity}
\vskip\baselineskip\hrule\smallskip\hrule\medskip
\halign to \hsize{\hfil#\hfil  &\hfil#\hfil &\hfil#\hfil &\hfil#\hfil 
                               &\hfil#\hfil ?? 
                               &\hfil#\hfil &\hfil#\hfil &\hfil#\hfil &\hfil#\hfil \cr
\noalign{\hskip 3.65cm \hbox{$\mu_{\ApJstyle e}=1$} \hskip 5.15cm
         \hbox{$\mu_{\ApJstyle e}=\mu_{\ApJstyle e}^{\ApJstyle\odot}
           =1.179$}}  
\noalign{\medskip\hrule\medskip}
Order &$\bar E_{\ApJstyle i}$ &$\kappa_{\ApJstyle i}$
      &$\xi_{\ApJstyle i}^{\ApJstyle\rm a}$
      &$\xi_{\ApJstyle i}^{\ApJstyle\rm s}$
      &$\bar E_{\ApJstyle i}$ &$\kappa_{\ApJstyle i}$
      &$\xi_{\ApJstyle i}^{\ApJstyle\rm a}$
      &$\xi_{\ApJstyle i}^{\ApJstyle\rm s}$ \cr
\noalign{\smallskip}
      &(MeV)
      &$\left({\rm cm^{\ApJstyle 2}\,g^{\ApJstyle -1}}\right)$
      &
      &
      &(MeV)
      &$\left({\rm cm^{\ApJstyle 2}\,g^{\ApJstyle -1}}\right)$
      &
      &  \cr
\noalign{\medskip\hrule\medskip}
\noalign{\vskip\baselineskip}
\noalign{\centerline{\rm$^{\ApJstyle 56}$Co}}
\noalign{\smallskip}
 0 &$ 1.24226$ &$0  .0643$ &$0  .7873$ &$0  .2127$ &$ 1.24226$ &$0  .0546$ &$0
  .7873$ &$0  .2127$\cr
 1 &$0 .23453$ &$0  .1542$ &$0  .3565$ &$0  .6435$ &$0 .23453$ &$0  .1308$ &$0
  .3565$ &$0  .6435$\cr
 2 &$0 .15085$ &$0  .1930$ &$0  .2518$ &$0  .7482$ &$0 .15085$ &$0  .1637$ &$0
  .2518$ &$0  .7482$\cr
 3 &$0 .11286$ &$0  .2206$ &$0  .1964$ &$0  .8036$ &$0 .11286$ &$0  .1870$ &$0
  .1964$ &$0  .8036$\cr
 4 &$0 .09070$ &$0  .2414$ &$0  .1616$ &$0  .8384$ &$0 .09070$ &$0  .2047$ &$0
  .1616$ &$0  .8384$\cr
 5 &$0 .07604$ &$0  .2577$ &$0  .1376$ &$0  .8624$ &$0 .07604$ &$0  .2185$ &$0
  .1376$ &$0  .8624$\cr
$\infty$ &0    &$0  .4006$ &0          &1          &0          &$0.  3397$ &0
       &1\cr
\noalign{\medskip}
\noalign{\centerline{\rm$^{\ApJstyle 56}$Ni}}
\noalign{\smallskip}
 0 &$0 .53479$ &$0  .0952$ &$0  .5911$ &$0  .4089$ &$0 .53479$ &$0  .0807$ &$0
  .5911$ &$0  .4089$\cr
 1 &$0 .18843$ &$0  .1701$ &$0  .3073$ &$0  .6927$ &$0 .18843$ &$0  .1443$ &$0
  .3073$ &$0  .6927$\cr
 2 &$0 .12885$ &$0  .2061$ &$0  .2235$ &$0  .7765$ &$0 .12885$ &$0  .1747$ &$0
  .2235$ &$0  .7765$\cr
 3 &$0 .09964$ &$0  .2311$ &$0  .1778$ &$0  .8222$ &$0 .09964$ &$0  .1960$ &$0
  .1778$ &$0  .8222$\cr
 4 &$0 .08177$ &$0  .2501$ &$0  .1484$ &$0  .8516$ &$0 .08177$ &$0  .2120$ &$0
  .1484$ &$0  .8516$\cr
 5 &$0 .06958$ &$0  .2650$ &$0  .1277$ &$0  .8723$ &$0 .06958$ &$0  .2247$ &$0
  .1277$ &$0  .8723$\cr
$\infty$ &0    &$0  .4006$ &0          &1          &0          &$0.  3397$ &0
       &1\cr
} 
\medskip\hrule
\vskip\baselineskip
     NOTE.---The solar $\mu_{\ApJstyle e}$ value was constructed using the solar system
abundances of Anders \&~Grevesse 1989.
\vfill
\supereject
}

{
\baselineskip=20pt

\baselineskip=20pt
\tabskip=50pt minus 50pt  
\newdimen\digitwidth
\setbox0=\hbox{\rm0}
\digitwidth=\wd0
\catcode\lq?=\active
\def?{\kern\digitwidth}

\centerline{TABLE II}
\centerline{Mean opacities and mean fractional component opacities}
\centerline{for solar and mean model W7 compositions}
\vskip\baselineskip\hrule\smallskip\hrule\medskip
\halign to \hsize{\hfil#\hfil  &\hfil#\hfil &\hfil#\hfil &\hfil#\hfil
                               &\hfil#\hfil ?? 
                               &\hfil#\hfil &\hfil#\hfil &\hfil#\hfil &\hfil#\hfil \cr
\noalign{
\hskip 2.775cm \hbox{Solar composition} \hskip 3.275cm 
          \hbox{Mean model W7 composition} }   
\noalign{\hskip 2.775cm
       ($\mu_{\ApJstyle e}=\mu_{\ApJstyle e}^{\ApJstyle\odot}
        =1.179$) \hskip 4.475cm
           ($\mu_{\ApJstyle e}=2.095$) }    
\noalign{\medskip\hrule\medskip}
Order &$\bar E_{\ApJstyle i}$ &$\kappa_{\ApJstyle i}$
      &$\xi_{\ApJstyle i}^{\ApJstyle\rm a}$
      &$\xi_{\ApJstyle i}^{\ApJstyle\rm s}$
      &$\bar E_{\ApJstyle i}$ &$\kappa_{\ApJstyle i}$
      &$\xi_{\ApJstyle i}^{\ApJstyle\rm a}$
      &$\xi_{\ApJstyle i}^{\ApJstyle\rm s}$ \cr
\noalign{\smallskip}
      &(MeV)
      &$\left({\rm cm^{\ApJstyle 2}\,g^{\ApJstyle -1}}\right)$
      &
      &
      &(MeV)
      &$\left({\rm cm^{\ApJstyle 2}\,g^{\ApJstyle -1}}\right)$
      &
      &  \cr
\noalign{\medskip\hrule\medskip}
\noalign{\vskip\baselineskip}
\noalign{\centerline{\rm$^{\ApJstyle 56}$Co}}
\noalign{\smallskip}
 0 &$ 1.24226$ &$0  .0547$ &$0  .7870$ &$0  .2130$ &$ 1.24226$ &$0  .0318$ &$0
  .7829$ &$0  .2171$\cr
 1 &$0 .23486$ &$0  .1307$ &$0  .3570$ &$0  .6430$ &$0 .23995$ &$0  .0827$ &$0
  .4393$ &$0  .5607$\cr
 2 &$0 .15099$ &$0  .1637$ &$0  .2524$ &$0  .7476$ &$0 .15294$ &$0  .1306$ &$0
  .4775$ &$0  .5225$\cr
 3 &$0 .11294$ &$0  .1873$ &$0  .1976$ &$0  .8024$ &$0 .11403$ &$0  .2006$ &$0
  .5810$ &$0  .4190$\cr
 4 &$0 .09074$ &$0  .2052$ &$0  .1639$ &$0  .8361$ &$0 .09147$ &$0  .3028$ &$0
  .6826$ &$0  .3174$\cr
 5 &$0 .07607$ &$0  .2194$ &$0  .1413$ &$0  .8587$ &$0 .07660$ &$0  .4459$ &$0
  .7630$ &$0  .2370$\cr
\noalign{\medskip}
\noalign{\centerline{\rm$^{\ApJstyle 56}$Ni}}
\noalign{\smallskip}
 0 &$0 .53479$ &$0  .0807$ &$0  .5911$ &$0  .4089$ &$0 .53479$ &$0  .0494$ &$0
  .6230$ &$0  .3770$\cr
 1 &$0 .18845$ &$0  .1443$ &$0  .3077$ &$0  .6923$ &$0 .19557$ &$0  .1043$ &$0
  .4762$ &$0  .5238$\cr
 2 &$0 .12888$ &$0  .1749$ &$0  .2245$ &$0  .7755$ &$0 .13679$ &$0  .1573$ &$0
  .5330$ &$0  .4670$\cr
 3 &$0 .09966$ &$0  .1964$ &$0  .1797$ &$0  .8203$ &$0 .10694$ &$0  .2298$ &$0
  .6205$ &$0  .3795$\cr
 4 &$0 .08180$ &$0  .2128$ &$0  .1515$ &$0  .8485$ &$0 .08797$ &$0  .3328$ &$0
  .7054$ &$0  .2946$\cr
 5 &$0 .06961$ &$0  .2259$ &$0  .1326$ &$0  .8674$ &$0 .07468$ &$0  .4764$ &$0
  .7756$ &$0  .2244$\cr
} 
\medskip\hrule
\vskip\baselineskip
     NOTE.---The solar composition is the solar system composition 
of Anders \&~Grevesse 1989.
The mean model~W7 composition is given by Thielemann~\etal 1986.

     The mean model~W7 composition used is the final composition
after all radioactive species have decayed.
Earlier time compositions give somewhat different mean opacities
from those in the table with the differences increasing with
scattering order.
The extreme case is the day~0 composition which
gives opacities that differ by $\sim 1\,$\% in the 0th order
and by up to $\sim 20\,$\% in the 5th order.
Experimentation, however, shows that the final composition
opacities can be used for all times without adding any significant
error in the energy deposition.
\vfill
\supereject
}

{

\baselineskip=20pt
\tabskip=50pt minus 50pt  
\newdimen\digitwidth
\setbox0=\hbox{\rm0}
\digitwidth=\wd0
\catcode\lq?=\active
\def?{\kern\digitwidth}

\centerline{TABLE III}
\centerline{Parameters for the radioactive decays}
\vskip\baselineskip\hrule\smallskip\hrule\medskip
\halign to \hsize{#\hfil    &#\hfil &#\hfil \cr
Parameter   &$^{\ApJstyle 56}$Co &$^{\ApJstyle 56}$Ni    \cr 
\noalign{\medskip\hrule\medskip}

$t_{\ApJstyle 1/2}$ (days)    &77.27    &5.9              \cr

$t_{\ApJstyle e}={t_{\ApJstyle 1/2}/\displaystyle \ln 2}$ (days)
                              &111.48   &8.5             \cr

$Q_{\ApJstyle\rm total}$ (MeV)
                              &4.5661  &2.136            \cr
\noalign{\vskip\baselineskip}

$Q_{\ApJstyle\gamma}$ (MeV)   &3.62(4) &1.72(2)          \cr

$C_{\ApJstyle\gamma}$ ($\rm ergs\,s^{\ApJstyle -1}\,g^{\ApJstyle -1}$)
                              &6.48(7)+9  &4.03(14)+10 \cr

$D_{\ApJstyle\gamma}$ ($\rm ergs\,s^{\ApJstyle -1}\,g^{\ApJstyle -1}$)
                              &7.02(8)+9  & ---           \cr
\noalign{\vskip\baselineskip}

$Q_{\ApJstyle\rm \Xray}$ (MeV)
                              &1.57(5)$-$3   &2.35(8)$-$3      \cr


$C_{\ApJstyle\rm \Xray}$ ($\rm ergs\,s^{\ApJstyle -1}\,g^{\ApJstyle -1}$)
                              &2.82(9)+6  &5.5(2)+7 \cr

$D_{\ApJstyle\rm \Xray}$ ($\rm ergs\,s^{\ApJstyle -1}\,g^{\ApJstyle -1}$)
                              &3.05(10)+6  & ---           \cr
\noalign{\vskip\baselineskip}

$Q_{\ApJstyle\beta^{\ApJstyle +}}^{\ApJstyle\rm KE}$ (MeV)
                              &0.116(6) &$\sim 0$         \cr

$Q_{\ApJstyle\beta^{\ApJstyle +}}^{\ApJstyle\rm KE}
 /Q_{\ApJstyle\gamma}$        &0.032(2) &$\sim 0$        \cr

$\beta^{\ApJstyle +}$ fraction (\%)
                              &19.0(9) &$<0.0013$  \cr

$\bar E_{\ApJstyle\beta^{\ApJstyle +}}^{\ApJstyle\rm KE}$
                              &0.61(3)  &$\sim 0$        \cr
 
$Q_{\ApJstyle\rm atomic\,\,el.}^{\ApJstyle\rm KE}$ (MeV) 
                              &3.6(3)$-$3 &6.9(3)$-$3       \cr

$C_{\ApJstyle\rm lepton}$
            ($\rm ergs\,s^{\ApJstyle -1}\,g^{\ApJstyle -1}$)
                              &2.14(11)+8 &1.62(9)+8     \cr

$D_{\ApJstyle\rm lepton}$
            ($\rm ergs\,s^{\ApJstyle -1}\,g^{\ApJstyle -1}$)
                              &2.32(12)+8 & ---          \cr
} 
\medskip\hrule
\vskip\baselineskip
     NOTE.---The meanings of most of the symbols follow from the
text.
The $t_{\ApJstyle 1/2}$ quantity is half-life.
The $\bar E_{\ApJstyle\beta^{\ApJstyle +}}^{\ApJstyle\rm KE}$ quantity
is the mean kinetic energy of a positron.
We have put the uncertainties in the last digits of the quantities
in brackets and have written $\times 10^{\ApJstyle\pm k}$ as
$\pm k$.

     The values have been taken or derived from Browne \&~Firestone
1986 and Huo 1992.
The uncertainties given in the references have been treated as
standard deviations in obtaining the uncertainties in the derived
quantities.
One of the quantities we derived was the
$^{\ApJstyle 56}$Co mean positron kinetic energy.
We calculated this assuming an allowed $\beta$-decay spectrum for all
three positron channels using the Fermi functions given by
Rose 1955.
The overwhelmingly dominant positron channel has a spectrum
that is consistent with an allowed $\beta$ decay (Pettersson,
Bergman, \&~Bergman 1965).

     The energy generation $C$ and $D$ coefficients 
for the positron and atomic electron kinetic energies
have been summed and subscripted by lepton.
We assume local deposition for both decay products,
and so they can be treated together.
The individual $C$ and $D$ coefficients for
positron and atomic electron kinetic energies can be obtained by scaling
the given $C$ and $D$ coefficients:  e.g., 
$C_{\ApJstyle\beta+}=C_{\ApJstyle\rm lepton}Q_{\ApJstyle\beta+}
                        /\left( Q_{\ApJstyle\beta+}
                               +Q_{\ApJstyle\rm atomic\,\,el.}\right)$.
\vfill
\supereject
}

}  

\specialpage
{\centerline{REFERENCES}
\baselineskip=20pt
\vskip\baselineskip
\reference

\refpaper Ambwani, K., \&~Sutherland, P. G. 1988, ApJ, 325, 820.

\refpaper Anders, E., \&~Grevesse, N.  1989, Geochim.~Cosmochim.~Acta,
              53, 197.  

\refbook Arfken, G. 1970, Mathematical Methods for Physicists
            (New York:  Academic Press).

\refindent Axelrod, T. S. 1980, Ph.D.~thesis, Univ.~of California,
             Santa~Cruz  

\refpaper Brown, B. L., \&~Leventhal, M. 1987, ApJ, 319, 637.

\refbook Browne, E., \&~Firestone, R. B. 1986, Table of Radioactive
          Isotopes (New York:  John Wiley \& Sons, Inc.).


\refpaper Chan, K. W., \&~Lingenfelter, R. E. 1993, ApJ, 405, 614.

\refpaper Cohen, E. R., \&~Taylor, B. N. 1987, J.~Research of the National
            Bureau of Standards, 92, 85.

\refpaper Colgate, S.~A., Petschek, A.~G., \&~Kriese, J.~T. 1980, ApJ, 237,
            L81.  

\refedited Davisson, C. M. 1965, in Alpha-, Beta-, and Gamma-Ray
             Spectroscopy,
             ed.~K.~Siegbahn
             (Amsterdam:  North-Holland), 37.


\refedited Fransson, C. 1994, in Supernovae:
             Session~LIV of the
             Les~Houches \'Ecole d'\'Et\'e de Physique Th\'eorique,
             ed.~S.~A.~Bludman, R.~Mochkovitch, \&~J.~Zinn-Justin
             (Amsterdam:  North-Holland), 677.


\refpaper G\'omez-Gomar, J., Isern, J., \&~Jean, P. 1998, MNRAS,
             295, 1, astro-ph/9709048.

\refindent Harkness, R. P. 1991, in {ESO/EIPC Workshop:  SN~1987A
             and Other Supernovae,} ed.~I.~J.~Danziger
              \&~K.~Kj\"ar (Garching:  ESO), 447



\refpaper H\"oflich, P., Khokhlov, A., \&~M\"uller, E. 1992, A\&A, 259,
            549.


\refpaper H\"oflich, P., Wheeler, J. C., \&~Khokhlov, A. 1998, ApJ, 492,
             228.

\refpaper Houck, J. C., \&~Fransson, C. 1996, ApJ, 456, 811.


\refindent Hubbell, J. H. 1969, NSRDS-NBS 29

\refpaper Huo, J. 1992, Nuclear Data Sheets, 67, 523.


\refindent Jeffery, D. J. 1998,
             in Stellar Evolution, Stellar Explosions, and Galactic
             Chemical Evolution:  Proc. 2nd Oak Ridge Symposium on
             Atomic \&~Nuclear
             Astrophysics, ed.~A.~Mezzacappa
             (Bristol:  Institute of Physics Publishing), 687,
             astro-ph/9802229




\refpaper Kumagai, S., Nomoto, K., Shigeyama, T., Hashimoto, M., \&~Itoh, M.
            1993, A\&A, 273, 153.

\refpaper Liu, W., Jeffery, D. J., \&~Schultz, D. R. 1997a,
             ApJ, 483, L107.

\refpaper \refrepeat 1997b,
             ApJ, 486, L35.

\refpaper Liu, W., Jeffery, D. J., Schultz, D. R., Quinet, P.,
             Shaw, J., \&~Pindzola, M. S. 1997c,
             ApJ, 489, L141.

\refpaper Liu, W., \&~Victor, G. A. 1994, ApJ, 435, 909.




\refbook Mihalas, D. 1978, Stellar Atmospheres (San Francisco:  Freeman).

\refpaper \refrepeat 1980a, ApJ, 237, 574.

\refpaper \refrepeat 1980b, ApJ, 238, 1034.

\refedited Milne, P. A., The, L.-S., Leising, M. D. 1997, in
             Proc.~The Fourth Compton Symposium, ed.~C.~D.~Dermer, 
             M.~S.~Strickman, \&~J.~D.~Kurfess (New York:
             American Institute of Physics Press), 1022.

\refedited Nomoto, K., Shigeyama, T., Kumagai, S., Yamaoka, H.,
             \&~Suzuki, T. 1994, in Supernovae:
             Session~LIV of the
             Les~Houches \'Ecole d'\'Et\'e de Physique Th\'eorique,
             ed.~S.~A.~Bludman, R.~Mochkovitch, \&~J.~Zinn-Justin
             (Amsterdam:  North-Holland), 489.

\refpaper Pettersson, H., Bergman, O., \&~Bergman, C. 1965, Arkiv f\"or
            Fysik, 29, 423.


\refedited Rose, M. E. 1955, in Beta- and Gamma-Ray Spectroscopy,
             ed.~K.~Siegbahn (New York:  Interscience Publishers Inc.),
             875.


\refindent Ruiz-Lapuente, P. 1997,
             in Proc. NATO ASI on
             Thermonuclear Supernovae, ed.~P.~Ruiz-Lapuente,
             R.~Canal, \&~J.~Isern
             (Dordrecht:  Kluwer), 681, astro-ph/9604094





\refpaper Ruiz-Lapuente, P., Lichti, G. G., Lehoucq, R., Canal, R., \&~Cass\'e, M.
            1993, ApJ, 417, 547.




\refindent Ruiz-Lapuente, P., \&~Spruit, H. C. 1998, ApJ, 500, 360,
             astro-ph/9711248

\refindent Sutherland, P. G. 1996, private communication

\refindent \refrepeat 1998, private communication

\refpaper Sutherland, P. G., \&~Wheeler, J. C. 1984, ApJ, 280, 282.

\refpaper Swartz, D. A., Sutherland, P. G., \&~Harkness, R. P. 1995,
            ApJ, 446, 766 (SSH).

\refpaper Thielemann, F.-K., Nomoto, K., \&~Yokoi, K. 1986, A\&A, 158, 17.

\refpaper Veigele, W. J. 1973, Atomic Data, 5, 51.

\refpaper Woosley, S. E. 1988, ApJ, 330, 218.







\refpaper Young, T. R., Baron, E., \&~Branch, D. 1995, ApJ, 449, L51.

}

\specialpage
\centerline{FIGURE LEGENDS}
\vskip\baselineskip

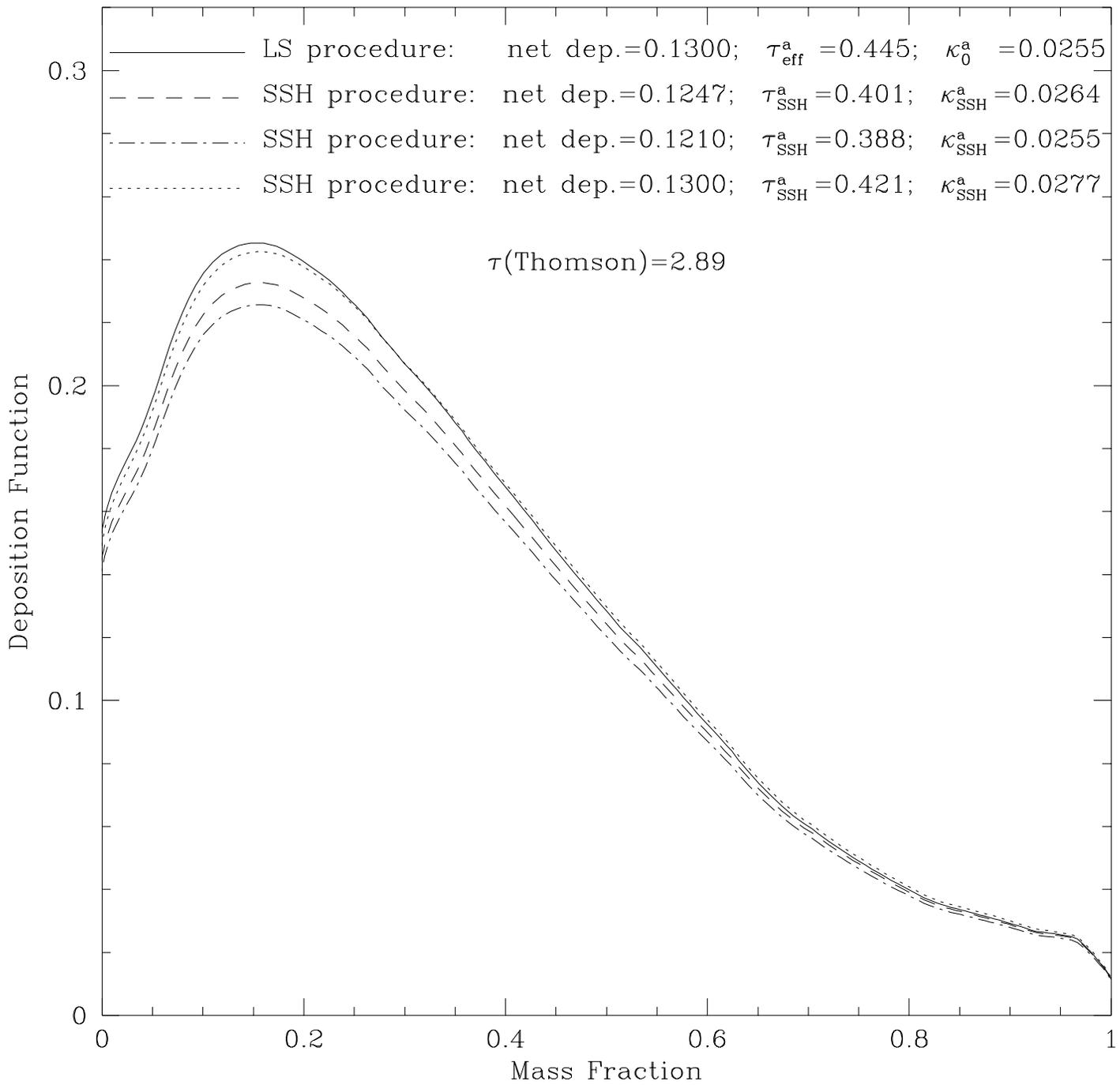
     FIG. 1.---The $\gamma$-ray energy deposition function for
model~W7 on day~110
calculated using the LS~procedure and the SSH~procedure with
three different values of the SSH mean opacity
$\kappa_{\ApJstyle\rm SSH}^{\ApJstyle\rm a}$.
The quantities in the table are described in the text.
The units of opacities in the table are
${\rm cm^{\ApJstyle 2}\,g^{\ApJstyle -1}}$.
The other quantities are dimensionless.
\bye